\documentstyle[12pt]{article}
\def\hybrid{\topmargin 0pt      \oddsidemargin 0pt
        \parskip 5pt plus 1pt   \jot = 1.5ex}
\catcode`\@=11
\def\marginnote#1{}

\newcount\hour
\newcount\minute
\newtoks\amorpm
\hour=\time\divide\hour by60
\minute=\time{\multiply\hour by60 \global\advance\minute by-\hour}
\edef\standardtime{{\ifnum\hour<12 \global\amorpm={am}%
        \else\global\amorpm={pm}\advance\hour by-12 \fi
        \ifnum\hour=0 \hour=12 \fi
        \number\hour:\ifnum\minute<10 0\fi\number\minute\the\amorpm}}
\edef\militarytime{\number\hour:\ifnum\minute<10 0\fi\number\minute}

\def\draftlabel#1{{\@bsphack\if@filesw {\let\thepage\relax
   \xdef\@gtempa{\write\@auxout{\string
      \newlabel{#1}{{\@currentlabel}{\thepage}}}}}\@gtempa
   \if@nobreak \ifvmode\nobreak\fi\fi\fi\@esphack}
        \gdef\@eqnlabel{#1}}
\def\@eqnlabel{}
\def\@vacuum{}
\def\draftmarginnote#1{\marginpar{\raggedright\scriptsize\tt#1}}

\def\draft{\oddsidemargin -0.1truein
        \def\@oddfoot{\sl preliminary draft \hfil
        \rm\thepage\hfil\sl\today\quad\militarytime}
        \let\@evenfoot\@oddfoot \overfullrule 3pt
        \let\label=\draftlabel
        \let\marginnote=\draftmarginnote
   \def\@eqnnum{{\rm (\theequation)}\rlap{\kern\marginparsep\tt\@eqnlabel}%
\global\let\@eqnlabel\@vacuum}  }
\newdimen\linethick  \linethick=0.4pt
\newdimen\hboxitspace    \hboxitspace=5pt
\newdimen\vboxitspace    \vboxitspace=5pt

\def\fr#1{%
\beq\new
\vcenter{
\hrule height\linethick
           \hbox{\vrule width\linethick
                 \kern\hboxitspace
                 \vbox{\kern\vboxitspace
                       \hbox{$\begin{array}{c}\displaystyle#1
          \end{array}$}%
                       \kern\vboxitspace}%
                 \kern\hboxitspace
                 \vrule width\linethick}%
           \hrule height\linethick}%
\eeq}
\newdimen\Squaresize \Squaresize=14pt
\newdimen\Thickness \Thickness=0.5pt

\def\Square#1{\hbox{\vrule width \Thickness
   \vbox to \Squaresize{\hrule height \Thickness\vss
      \hbox to \Squaresize{\hss#1\hss}
   \vss\hrule height\Thickness}
\unskip\vrule width \Thickness}
\kern-\Thickness}

\def\Vsquare#1{\vbox{\Square{$#1$}}\kern-\Thickness}

\def\numberbysection{\@addtoreset{equation}{section}
        \def\theequation{\thesection.\arabic{equation}}}
\numberbysection

\renewcommand{\theequation}{\thesection.\arabic{equation}}
\newcommand{\l@qq}[2]{\addvspace{2em}
 \hbox to\textwidth{\hspace{1em}\bf #1 \dotfill #2}}

\newcounter{app}

\def\app{\setcounter{equation}{0}
\def\theequation{\Alph{app}.\arabic{equation}}\par
   \addvspace{4ex}
   \@afterindentfalse
  \secdef\@app\@dapp}
\newcommand\@app{\@startsection {app}{1}{0ex}%
                                   {-3.5ex \@plus -1ex \@minus -.2ex}%
                                   {2.3ex \@plus.2ex}%
                                   {\normalfont\Large\bf}}

\def\@dapp#1{%
{\parindent \z@ \raggedright  \bf #1}\par\nobreak}
\def\l@app#1#2{\ifnum \c@tocdepth >\z@
    \addpenalty\@secpenalty
    \addvspace{1.0em \@plus\p@}%
    \setlength\@tempdima{2.5em}%
    \begingroup
      \parindent \z@ \rightskip \@pnumwidth
      \parfillskip -\@pnumwidth
      \leavevmode \bfseries
      \advance\leftskip\@tempdima
      \hskip -\leftskip
      #1\nobreak\hfil \nobreak\hb@xt@\@pnumwidth{\hss #2}\par
    \endgroup\fi}
\newcounter{sapp}[app]

\def\sapp{\def\theequation{\Alph{app}.\arabic{equation}}\par
   \@afterindentfalse
  \secdef\@sapp\@dsapp}
\newcommand\@sapp{\@startsection{sapp}{2}{\z@}%
                                     {-3.25ex\@plus -1ex \@minus -.2ex}%
                                     {1.5ex \@plus .2ex}%
                                     {\normalfont\large\bfseries}}

\def\@dsapp#1{%
{\parindent \z@ \raggedright  \bf #1}\par\nobreak}
\newcommand{\l@sapp}{\@dottedtocline{2}{1.5em}{3em}}

\def\titlepage{\@restonecolfalse\if@twocolumn\@restonecoltrue\onecolumn
     \else \newpage \fi \thispagestyle{empty}\c@page\z@
        \def\thefootnote{\fnsymbol{footnote}} }

\def\endtitlepage{\if@restonecol\twocolumn \else  \fi
        \def\thefootnote{\arabic{footnote}}
        \setcounter{footnote}{0}}  %\c@footnote\z@ }
\relax

\hybrid

\newtoks\@stequation

\def\subequations{\refstepcounter{equation}%
  \edef\@savedequation{\the\c@equation}%
  \@stequation=\expandafter{\theequation}%   %only want \theequation
  \edef\@savedtheequation{\the\@stequation}% %expanded once
  \edef\oldtheequation{\theequation}%
  \setcounter{equation}{0}%
  \def\theequation{\oldtheequation\alph{equation}}}

\def\endsubequations{%
  \setcounter{equation}{\@savedequation}%
  \@stequation=\expandafter{\@savedtheequation}%
  \edef\theequation{\the\@stequation}%
  \global\@ignoretrue}

\makeatletter
\newdimen\normalarrayskip              % skip between lines
\newdimen\minarrayskip                 % minimal skip between lines
\normalarrayskip\baselineskip
\minarrayskip\jot
\newif\ifold             \oldtrue            \def\new{\oldfalse}
\def\arraymode{\ifold\relax\else\displaystyle\fi} % mode of array enrties
\def\eqnumphantom{\phantom{(\theequation)}}     % right phantom in eqnarray
\def\@arrayskip{\ifold\baselineskip\z@\lineskip\z@
     \else
     \baselineskip\minarrayskip\lineskip1\baselineskip\fi}

\def\@arrayclassz{\ifcase \@lastchclass \@acolampacol \or
\@ampacol \or \or \or \@addamp \or
   \@acolampacol \or \@firstampfalse \@acol \fi
\edef\@preamble{\@preamble
  \ifcase \@chnum
     \hfil$\relax\arraymode\@sharp$\hfil
     \or $\relax\arraymode\@sharp$\hfil
     \or \hfil$\relax\arraymode\@sharp$\fi}}

\def\@array[#1]#2{\setbox\@arstrutbox=\hbox{\vrule
     height\arraystretch \ht\strutbox
     depth\arraystretch \dp\strutbox
     width\z@}\@mkpream{#2}\edef\@preamble{\halign \noexpand\@halignto
\bgroup \tabskip\z@ \@arstrut \@preamble \tabskip\z@ \cr}%
\let\@startpbox\@@startpbox \let\@endpbox\@@endpbox
  \if #1t\vtop \else \if#1b\vbox \else \vcenter \fi\fi
  \bgroup \let\par\relax
  \let\@sharp##\let\protect\relax
  \@arrayskip\@preamble}
\def\eqnarray{\stepcounter{equation}%
              \let\@currentlabel=\theequation
              \global\@eqnswtrue
              \global\@eqcnt\z@
              \tabskip\@centering                      %formulae centering
              \let\\=\@eqncr
              $$%
            \halign to \displaywidth  \bgroup
             \eqnumphantom \@eqnsel
      \hskip\@centering                               %right tab%
    $\displaystyle  \tabskip\z@ {##}$%
    &\global\@eqcnt\@ne \hskip 2\arraycolsep
         $ \displaystyle  \arraymode{##}$\hfil
    &\global\@eqcnt\tw@ \hskip 2\arraycolsep
         $\displaystyle\tabskip\z@{##}$\hfil
         \tabskip\@centering
    &{##}\tabskip\z@\cr}
\makeatother
\def\bea{\begin{eqnarray}}
\def\eea{\end{eqnarray}}

\def\beq{\begin{equation}}
\def\eeq{\end{equation}}
\def\be{\beq\new\begin{array}{c}}
\def\ee{\end{array}\eeq}
\def\bse{\begin{subequations}}                %%%SUBEQUATIONS
\def\ese{\end{subequations}}                 %
\begin{document}
\vspace{0.2cm}
\begin{center}
{\LARGE \bf
Nonabelian Duality and Solvable } \\
\vspace{0.5cm}{\LARGE \bf Large $N$ Lattice Systems. } \\
\vspace{0.7cm} {\large\bf Andrey Yu. Dubin}\\
\vspace{0.5cm}{\bf ITEP, B.Cheremushkinskaya 25, Moscow 117259, Russia}\\
\vspace{0.5cm}{{\it e-mail: dubin@vxitep.itep.ru}}
 \end{center}
\vspace{0.6cm}

\begin{abstract}

We introduce the basics of the nonabelian duality transformation of
$SU(N)$ or $U(N)$ vector-field models defined on a lattice.
The dual degrees of freedom are certain species of the integer-valued fields
complemented by the symmetric groups' $\otimes_{n} S(n)$ variables. While
the former parametrize relevant irreducible representations, the latter play
the role of the Lagrange multipliers facilitating the fusion rules involved.
As an application, I construct a novel solvable family of $SU(N)$ $D$-matrix
systems graded by the rank $1\leq{k}\leq{(D-1)}$ of the manifest
$[U(N)]^{\oplus k}$ conjugation-symmetry. Their large $N$ solvability is due to a
hidden invariance (explicit in the dual formulation) which allows for a
mapping onto the recently proposed eigenvalue-models \cite{Dub1} with the
largest $k=D$ symmetry. Extending \cite{Dub1}, we reconstruct a
$D$-dimensional gauge
theory with the large $N$ free energy given (modulo the volume factor)
by the free energy of a given proposed $1\leq{k}\leq{(D-1)}$ $D$-matrix
system. It is emphasized that the developed formalism provides
with the basis for higher-dimensional generalizations of the Gross-Taylor
stringy representation of strongly coupled $2d$ gauge theories.

\end{abstract}

\begin{center}
\vspace{0.5cm}{Keywords: Lattice, Yang-Mills, Duality, Solvability} \\
\vspace{0.2cm}{PACS codes 11.15.Ha; 11.15.Pg; 11.15.Tk}
\end{center}

\newpage

\section{Introduction}

A Duality of the $D=4$ continuum Yang-Mills gauge system to a kind of
string theory remains to be one of a few intuitive guiding principles
to attack nonperturbative dynamics of the strong interactions. Among
the circumstantial evidences, the central role is played by
the Wilson's $D\geq{2}$ string-like representation \cite{Wils} of the
strong-coupling (SC) series but in a $lattice$ cousin of the continuum YM
theory. As it is well known, this
particular expansion (running in terms of the inverse powers of the bare
coupling constant) can not be directly extended into the weak-coupling phase
relevant for the continuum limit. Nevertheless, we believe that (properly
chosen) lattice $YM$ systems hide a stringy pattern relevant for the
$D\geq{2}$ continuum gauge theories at least in the regime being the {\it
continuum} counterpart of the lattice SC phase. To this aim, one is to
consider a continuum $D\geq{2}$ $YM_{D}$ model with a {\it finite}
ultra-violate cut off $\tilde{\Lambda}_{UV}$ and sufficiently {\it large}
coupling constant(s).

To support this idea, we refer to the well-studied $D=2$ case, where
Gross and Taylor proposed an elegant stringy representation \cite{Gr&Tayl} of
the large $N$ SC series in the continuum $SU(N)$ gauge system on an arbitrary
$2d$ surface. Recall that a continuum $YM_{2}$ can be directly reproduced
through the corresponding lattice gauge model with the action defined via the
associated self-reproducing plaquette-factor \cite{Migd75,Witt91}. As a
result, the pattern of the
proposed $D=2$ representation is the $same$ both in the SC regime of a given
continuum $YM_{2}$ theory and in the SC phase of
the corresponding self-reproducing lattice model.
The only considerable distinction is that in the latter case one would deal
with the discretized surfaces rather than with the $2d$ manifolds.
The crucial point is that, as far as the 'built in' topological data is
concerned, this difference
does {\it not} matter. As a result, in both instances the appropriate SC series can
be reinterpreted in terms of statistics of all admissible branched coverings
(associated to the base-surface) described canonically in terms of the
symmetric groups' elements.

The key-ingredient of the above $D=2$ construction is the
so-called Schur-Weyl complementarity (see e.g. \cite{Moore} for a review)
between
the Lie and symmetric groups. Altogether, for $YM_{2}$ it fulfils the role of a
bridge between the symmetry and topology that makes it particularly suitable
for construction of the gauge string representation. Unfortunately, the proposed
in \cite{Gr&Tayl} technology can not be directly extended to $D\geq{3}$.
The purpose of the present paper is to develope the basics of an approach
which, among other things, renders accessible higher-dimensional
generalizations of the Gross-Taylor pattern.

One of the central elements of our approach is the nonabelian duality
transformation. On the one hand, the
latter can be considered as a natural extension of the Schur-Weyl duality. On
the other hand, it is to be viewed as a realization of the long sought
nonabelian version of the abelian transformation well known in the context of
the pure $U(1)$ lattice system (see \cite{Sav} for a review). This
generalization  can be compared, in particualar, with the recent conjecture of
Polyakov \cite{Polyak}. He advocated that the lattice abelian transformation
encodes the $N=1$ string-like pattern which might be generalized (in a yet
unknown way) for the $U(N)$ continuum gauge theory with an arbitrary $N$.

The Gauge String construction, we keep in mind, is facilitated by the
formalism which is to synthesize both the nonabelian duality (i.e.
{\it symmetry}) and the {\it topology} of the branched coverings.
We find it appropriate to introduce the former ingredient in a simpler
setting that avoids entanglement with the topology. For this purpose, we
find a simpler application of the duality transformation constructing a
family of solvable large $N$ $SU(N)$ matrix systems which are $not$ tractable
by other methods. The $D>2$ generalization of the Gross-Taylor $2d$ stringy
pattern on a lattice will be given in a forthcoming paper \cite{Dub2}.

In what follows, employing the nonabelian duality we design a mapping
between the recently proposed solvable $D$-matrix eigenvalue-theories
\cite{Dub1} and a novel class of the $D$-matrix models which apparently are
{\it not} of the eigenvalue-type. There are a few reasons why the latter
models are worth studying by themselves. First, it provides with a rare
example of solvable {\it multi}matrix models nontrivially depending on the
nondiagonal components of the $SU(N)$ or $U(N)$ matrices involved.
As we will see, the mechanism behind their solvability is different from that
in the popular systems computable owing to the 'built in' Itzykson-Zuber
integral \cite{Itz&Zub}:
the Kazakov-Migdal model \cite{Kaz&Mig} together with the conventional 
and conformal multimatrix systems (see e.g. \cite{Moroz} for a review).
Second, generalizing the prescription of \cite{Dub1}, the new models
can be viewed as the large $N$ reduction of the associated $D$-dimensional
lattice gauge theories. In other words, the large $N$ partition function
(PF) $\tilde{X}_{L^{D}}$ of the latter theory (defined on a cubic lattice of
$D$-volume $L^{D}$) can be reproduced
\be
\lim_{N\rightarrow{\infty}} \tilde{X}_{L^{D}}=
\lim_{N\rightarrow{\infty}}
(\tilde{X}_{r})^{L^{D}}~,
\label{2.10}
\ee
through the PF $\tilde{X}_{r}$ of the corresponding $D$-matrix
model with the $reduced$ space-time dependence. The latter models are
computable owing to the claimed duality to the
basic $SU(N)$ eigenvalue-family \cite{Dub1}
\be
e^{-S^{(2)}_{r}(\{U_{\rho}\})}=\sum_{\{R_{\mu\nu}\}}
e^{-S^{(2)}(\{R_{\mu\nu}\})}
\prod_{\mu\nu=1}^{D(D-1)/2}
|\chi_{R_{\mu\nu}}(U_{\mu})\chi_{R_{\mu\nu}}(U_{\nu})|^{2},
\label{1.35}
\ee
formulated in terms of the {\it eigenvalues} of the $D$-matrices
$U_{\rho}\in{SU(N)}$,
or to its minor modification
\be
e^{-S^{(1)}_{r}(\{U_{\rho}\})}=
\sum_{\{R_{\phi}\}}e^{-S^{(1)}(\{R_{\phi}\})}
\prod_{\{\mu\nu\}} 
\chi_{R_{\mu\nu}}(U^{+}_{\mu})\chi_{R_{\mu\nu}}(U^{+}_{\nu})
\prod_{\rho=1}^{D} \chi_{R_{\rho}}(U_{\rho}),
\label{1.36}
\ee
both being defined on a single $D$-cube with periodic boundary conditions:
$\{\mu\nu\}=\{1,...,D(D-1)/2\}$. The relevant sums run over all $SU(N)$
irreducible representations (irreps) $R_{\phi}\in{Y_{n(\phi)}^{(N)}},~\phi
\in{\{\mu\nu\},
\{\rho\}}$, and in the case of (\ref{1.36}) it is postulated that the numbers of boxes
$n(\rho)$ in the associated Young tableau are constrained by $n(\rho)=
\sum_{\nu\neq{\rho}}^{D-1} n(\rho\nu)$.

The key-advantage of the eigenvalue-systems
(\ref{1.35}),(\ref{1.36}) is that their large $N$ PF
$\tilde{X}^{(m)}_{r}$ is explicitly computed \cite{Dub1} employing the
saddle-point (SP) method applied to the irreps $\{R_{\phi}\}$. To construct
the purported solvable noneigenvalue deformations of
(\ref{1.35}),(\ref{1.36}), we first rewrite
the PF $\tilde{X}^{(m)}_{r}$ of the latter systems in terms \cite{Dub1} of
the $D$-products
\be
\lim_{N\rightarrow{\infty}} \tilde{X}^{(m)}_{r}=\lim_{N\rightarrow{\infty}}
[~\sum_{\{R_{\phi}\}}
e^{-\frac{S(\{R_{\phi}\})}{m}}
\otimes_{\rho=1}^{D} L^{(D-1)}_{R_{\rho}|\{R_{\rho\nu}\}}~]^{m}
\label{6.3p}
\ee
of the generalized Littlewood-Richardson (GLR) coefficients of $(D-1)th$
order
\be
L^{(D-1)}_{R_{\rho}|\{R_{\rho\nu}\}}=
\int d\tilde{U}^{SU(N)}_{\rho}~\chi_{R_{\rho}}(\tilde{U}^{+}_{\rho})~
[\otimes_{\nu\neq{\rho}}^{D-1}\chi_{R_{\rho\nu}}(\tilde{U}_{\rho})]~\in{~\bf
Z_{\geq{0}}}~.
\label{6.3}
\ee
which encode the fusion rules of the $SU(N)$ characters.
I assert that {\it the
reduction of the PF to the GLR {\it generating functional} 
(\ref{6.3p}) takes place in a
larger, apparently {\it non}eigenvalue variety
of $D$-matrix models} like
\be
e^{-S_{r}(\{U_{\rho}\})}=\sum_{\tilde{n}(+)\in{\bf Z_{\geq{0}}}}
\sum_{\{w^{(s)}_{q}\}_{\tilde{n}(+)}} e^{-A(\{w^{(s)}_{q}\})}Re[
\otimes_{q=1}^{p} tr(U_{\mu_{q}}U_{\nu_{q}}... U_{\lambda_{q}})],
\label{2.0}
\ee
where the $\{w^{(s)}_{q}\}_{\tilde{n}(+)}$-sum runs over all
cyclic-symmetrized $words$ $w^{(s)}_{q}\equiv{[\mu_{q}\nu_{q}...\lambda_{q}}]^{(s)}$
(made of the $2\otimes D$ different $\rho$-, ${\rho}^{-1}$-'letters')
of lengths $n_{k}$ which can be composed from the total number
$\tilde{n}(+)$ of the $\{U_{\rho^{\pm}}\}$ factors:
$\sum_{q=1}^{p} n_{q}=\tilde{n}(+)$.

The representation (\ref{2.0}) is not particularly helpful to distinguish the
GLR computable variety, we are interested in, from the generic
$D$-matrix system. The only explicit
general structure is that the family (\ref{2.0}) forms a natural
hierarchy graded by the rank $k=D,...,1,$ of the $[U(N)]^{\oplus k}$
conjugation-invariance
\be
[U(N)]^{\oplus k}:~~
U_{\rho({\beta})}\rightarrow{g^{+}_{\beta}~
U_{\rho({\beta})}~g_{\beta}}
~~,~~{\beta}=1,...,k~~,~~g_{\beta}\in{U(N)},
\label{2.0d}
\ee
where $U_{\rho}$ factors are recollected into a set of $\beta$-families
$\{\rho({\beta})\}$. The deep reason for the GLR solvability of the $k<D$
{\it non}eigenvalue-systems (i.e. for their duality to the $k=D$ models
(\ref{1.35}),(\ref{1.36})) is a specific $hidden$ symmetry. The latter
becomes manifest only in the dual representation for the PF of
a $k\leq{D}$ subvariety of (\ref{2.0}) (including (\ref{1.35}),
(\ref{1.36})). Extending the abelian transformation \cite{Sav}, we propose
the following dual variables: the $\phi$-species of the
integer-valued
$\{\lambda(\phi)\}$-sets (parametrizing associated $SU(N)$ irreps $R_{\phi}$)
combined with the elements of the symmetric groups $\bigotimes_{n} S(n)$.
The $S(n)$-valued degrees of freedom (being composed into the
$characters$ of the corresponding tensor representations) act as the
{\it Lagrange multipliers}. They facilitate the nonabelian fusion-rule
constraints for the complementary integer-valued fields
$\lambda_{i}(\phi),~i=1,...,N-1,$ entering the construction within the
canonical $S(n_{\phi})$-valued Young projectors $P_{R_{\phi}},~R_{\phi}
\in{Y_{n_{\phi}}^{(N)}}$.

In the dual representation, the GLR coefficients in the PF (\ref{6.3p})
can be refered (owing to the Schur-Weyl complementarity \cite{Gr-in-phys}) to
the $simplest$ available fusion-pattern of the $S(n_{\phi})$-valued Young
idempotents (YI) $C_{R_{\phi}}\sim{P_{R_{\phi}}}$. It is the nonabelian
duality which, as we will see, allows to reveal that among (\ref{2.0}) there
are $k<D$ systems with the same (as in the $k=D$ case (\ref{1.35}),
(\ref{1.36})) GLR pattern (\ref{6.3p}) of the underlying YI fusion-rules.
Complementary, the pivotal role of the $\bigotimes_{n} S(n)$-variables in
fact foreshadows a tight relation to the $D\geq{2}$
Gauge String construction generalizing the $2d$ pattern \cite{Gr&Tayl}.

Finally, the organization of the paper is as following.
The details of the {\it exact} duality transformation, applied first to the
eigenvalue-models (\ref{1.35}),(\ref{1.36}), are discussed at length in
in Section 2. In particular, for the latter models we rederive the GLR
form (\ref{6.3p}) of the PF  $\tilde{X}^{(m)}_{r}$ directly in the
framework of the dual representation. Building on this formalism, in Section 3
we construct the GLR computable $k\leq{D-1}$ subvariety of the $D$-matrix
systems (\ref{2.0}). Among the deformations, we select a $k=1$
{\it non}eigenvalue-family specifically suitable for the discussion of the
continuum limit (CL) in the corresponding $D$-dimensional
induced gauge theories. The algorithm to reconstruct the latter
theories is formulated in Section 4. To address the issue of the CL, in
Section 5 we first transform  the associated large $N$ GLR functional
(\ref{6.3p}) into the 1-matrix representation \cite{Dub1}. Then
we prove that the simple criterion imposed on the latter 1-matrix system
(formulated in \cite{Dub1} for the $k=D$ family (\ref{1.35}),(\ref{1.36})) is
valid in
the case of the selected $k=1$ variety as well. For this purpose, following
\cite{Dub1} we demonstrate that the link-variables in the corresponding
induced lattice gauge theory are localized
$\{U_{\rho}({\bf z})\rightarrow{\hat{1}}\}$ (modulo the relevant symmetries)
which is tantamount to the regime of the CL.

Our conclusion emphasizes the major novel possibilities open by the proposed approach.
Among other things, we assert one of the expected features of the
$D>{2}$ {\it Gauge~String} representation (of the $strongly$ coupled $YM_{D}$
theories) novel compared to the $D=2$ construction of Gross and Taylor
\cite{Gr&Tayl}. A few Appendices contain relevant technical details of the
derivations used in the main text.

\section{The Dual form of $D$-matrix models.}
\setcounter{equation}{0}

Let us introduce the concept of the nonabelian Duality 
which is built on the $complementarity$ of the Lie and symmetric groups
(reviewed in Appendix A). In our opinion, the dual representation provides
with the appropriate
mathematical framework to operate with the results of the generic multiple
$U(N)$ or $SU(N)$ integrations like those defining the PF $\tilde{X}_{r}$
of an arbitrary $D$-matrix model (\ref{2.0}).

It is instructive to view the nonabelian transformation as the extension
of the abelian construction \cite{Sav} dating back to the
classical paper due to Kramers and Wannier who discovered the selfduality of
the $2d$ Ising model. For a brief review, take the most relevant for our
analysis option of the pure lattice gauge theory based on the $U(1)$ group,
i.e. compact QED in $D\geq{2}$. Its partition function is defined as the
multiple link-integral of the product composed of the standard $U(1)$
plaquette-factors $Z(U)=\sum_{n\in{\bf Z}} \chi_{n}(U)~Q(n)$
\be
\tilde{X}^{U(1)}_{N_{p},N_{l}}=\prod_{p=1}^{N_{p}}\prod_{l=1}^{N_{l}}
\int dU^{U(1)}_{l}~Z(U(p))~~~~~,~~~~~
\int dU^{U(1)}_{l}\equiv{\int_{-\pi}^{\pi}} \frac{d\theta_{l}}{2\pi}~,
\label{2.60}
\ee
where $U^{U(1)}_{l}=e^{i\theta_{l}}$, while
$U(p)=e^{i[\bigtriangledown{\theta}](p)}$ is the holonomy around the $p$th
elementary plaquette of the base-lattice $\{p;l\}$ consisting of $N_{l}$
links and $N_{p}$ plaquettes. In the abelian case, the crucial simplification
arises due to the fact that
all $U(1)$ irreps (labelled by $n\in{\bf Z}$) are $one$-dimensional, while
the $U(1)$ characters
\be
U(1):~~\chi_{n}(U)=U^{\oplus n}=e^{i\theta n}~~~,~~~
\chi_{n_{1}}(U)\chi_{n_{2}}(U)=\chi_{n_{1}+n_{2}}(U)~,
\label{2.61}
\ee
form the so-called {\it character-group} \cite{Sav} isomorphic to
${\bf Z}$. Combining eq. (\ref{2.61}) with the simple structure the generic
$U(1)$ 1-link integral
\be
\int dU^{U(1)}~U^{\oplus n}~(U^{+})^{\oplus m}=\delta[n,m]~~~,~~~
n,m\in{\bf Z_{\geq{0}}}~,
\label{2.62}
\ee
one easily derives the dual form of the partition function (\ref{2.60}) as
the constrained multiple sum over the {\it integer-valued} variables (assigned to
the plaquettes)
\be
\tilde{X}^{U(1)}_{N_{p},N_{l}}=\prod_{p=1}^{N_{p}} \sum_{n(p)\in{\bf Z}}
Q(n(p))~\prod_{l=1}^{N_{l}}~
\delta[(\sum_{\tilde{p}_{l}=1}^{2(D-1)} n(\tilde{p}_{l})),0]~.
\label{2.63}
\ee
Here the sum in the argument of the Kronecker delta-function, running over the
$2(D-1)$ plaquettes $\tilde{p}_{l}$ which share a given link $l$ in common,
represents the relevant $U(1)$ fusion-rule algebra.

Finally, one observes that the relevant $U(1)$ (and more generally
$U(N)$ or $SU(N)$) $D$-matrix models (\ref{2.0})
are defined on the following $reduced$ base-lattice. Topologically, this
lattice
is made (as it is clear e.g. from the pattern of eq. (\ref{1.35})) 
from the set of the $D(D-1)/2$ distinct $\mu\nu$-plaquettes through the
identification of all $D-1$ their $\rho$-links to match with the topology of
a $D$-cube with periodic boundary conditions.
Altogether this conglomerate can be visualized as 
$D(D-1)/2$ mutually intertwined 2-tora.

\subsection{The Dual form of the $U(N)$ measure.}

Let us proceed introducing the dual representation of the functional measure
in the nonabelian lattice vector-field theories. Recall that the measure,
considered
as a $distribution$, can be defined specifying all the '$moments$' of this
distribution. On a given base-lattice, these 'moments' are specified defining
at each link the set of generic
1-link integrals $M^{G}(n,m)_{j_{1}...l_{m}}^{p_{1}...q_{m}}$
\be
\int {dU} (U)^{p_{1}}_{j_{1}}...(U)^{p_{n}}_{j_{n}}~
(U^{+})^{q_{1}}_{l_{1}}...  (U^{+})^{q_{m}}_{l_{m}}
\equiv{\int{dU}D(U)_{\{j^{\oplus n}\}}^{\{p^{\oplus n}\}}
D(U^{+})_{\{l^{\oplus m}\}}^{\{q^{\oplus m}\}}}
\label{3.5}
\ee
composed from the $N\times N$ matrices $(U)^{p_{k}}_{j_{k}},~
(U^{+})^{q_{k}}_{l_{k}}$ in the (anti)fundamental
representation of the Lie group $G$ in question. The crucial observation is
that $M^{G}(n,m)$ can be dually reformulated in
terms of the $S(n)$-valued variables. In particular, in the $G=U(N)$ case
$M^{U(N)}(n,m)$ reads
\be
M^{U(N)}(n,m)_{j_{1}...l_{m}}^{p_{1}...q_{m}}=\delta[n,m]
\sum_{\delta\in{S(n)}}{D(\delta^{-1} \Lambda^{(-1)}_{n})}
^{\{q^{\oplus n}\}}_{\{j^{\oplus n}\}}~
D(\delta)^{\{p^{\oplus n}\}}_{\{l^{\oplus n}\}}
\label{3.6}
\ee
generalizing the $U(1)$ pattern (\ref{2.62}). The derivation of the
identity (\ref{3.6})
is given in Appendix B, and here we simply explain the meaning of its
building blocks. The factor $D(\Psi)$ stands for the
canonical tensor representation of a given $S(n)$-algebra element $\Psi$
deduced (by linearity) from the representation \cite{Gr-in-phys} of a
$S(n)$-group element $\sigma$
\be
D(\sigma)_{\{j^{\oplus n}\}}^{\{i^{\oplus n}\}}=
\delta_{j_{1}}^{i_{\sigma(1)}}
\delta_{j_{2}}^{i_{\sigma(2)}}...\delta_{j_{n}}^{i_{\sigma(n)}}~~~~;~~~~
\hat{\sigma}:~k\rightarrow{\sigma(k)}~,~k=1,...,n,
\label{3.7}
\ee
while the introduced in (\ref{3.6}) operator $\Lambda^{(-1)}_{n}\in{S(n)}$
can be viwed as belonging to the family
\be
\Lambda^{(m)}_{n}=
\sum_{R\in{Y_n^{(N)}}} d_{R}~(n!~dimR/d_{R})^{m}~C_{R}~~~,~~~m\in{\bf Z}~,
\label{3.8}
\ee
which is expressed in terms of the dimensions $d_{R}$ and $dimR$ of
the $S(n)$- and (chiral) $U(N)$-irreps $R\in{Y_n^{(N)}}$ respectively. As for
the operator $C_{R}\in{S(n)}$ in eq. (\ref{3.8}), it denotes the canonical
Young idempotent proportional $P_{R}=d_{R}C_{R}$ to the Young projector
$P_{R}$ \cite{Gr-in-phys} (see Appendix A for more details) 
\be
P_{R}={\frac{d_{R}}{n!}}\sum_{\sigma\in S(n)}\chi_{R}(\sigma)~\sigma~~,~~
R\in{Y_{n}}~,
\label{3.9}
\ee
where $\chi_{R}(\sigma)$ is the corresponding character.
Summarizing, one observes that the $S(n)$-representation (\ref{3.6}) of the
1-link integral (\ref{3.5}) establishes a remarkable $duality$ extending the
Schur-Weyl complementarity (see Appendix A for a review of the latter).
Considered as
the operators acting on $|p>^{\oplus{n}}\otimes |q>^{\oplus{n}}$-space, the
left and
right hand sides of eq. (\ref{3.6}) belong to the complementary structures:
the Lie group $ring$ and the symmetric group $algebra$ (being augmented by
the integer-valued fields parametrizing irreps $R$) respectively.

Next, in contradistinction to the $U(N)$ case, the $SU(N)$ 1-link integral
(\ref{3.5}) doesn't vanish provided that
$n=m \bmod N$ \cite{Dr&Zub} which makes it generically more complex.
In the context of the $D$-matrix systems (\ref{2.0}), the important
simplification arises because (without loss of generality)
the $total$ amounts $n_{\pm}(\rho)$ of the $U_{\rho^{\pm}}$ factors in each
trace-product of (\ref{2.0}) can be constrained to be $equal$ for each $\rho$.
In other words, given the numbers $n_{\pm}^{(q)}(\rho)$ of the
$U_{{\rho}^{\pm}}$ factors entering corresponding traces
$tr(U_{\mu_{q}}U_{\nu_{q}}... U_{\lambda_{q}})$,
we postulate that
\be
n_{+}(\rho)=n_{-}(\rho)\equiv{n(\rho)}
~~,~~\forall{\rho}~~~;~~~
n_{\pm}(\rho)=\sum_{q=1}^{p} n_{\pm}^{(q)}(\rho).
\label{6.20v}
\ee
In turn, the constraint (\ref{6.20v}) ensures that the action in the
associated $D$-matrix $SU(N)$ subfamily of 
(\ref{2.0}) is invariant under the $D$ copies of the transformations
\be
[U(1)]^{\oplus D}:~U_{\rho}\rightarrow{t_{\rho}U_{\rho}}~~,~~
t_{\rho}\in{U(1)}~,
\label{2.9}
\ee
taking values in $U(1)$ rather than in the center-subgroup $T={\bf Z_{N}}$ of
$SU(N)$. As a result, the nondiagonal moments $M^{SU(N)}(n,m)$, $n\neq{m}$,
do $not$ contribute into the PF $\tilde{X}_{r}$, while in the remaining
diagonal integrals $M^{SU(N)}(n,n)$ the $SU(N)$ link-variables can be
substituted by the $U(N)=[SU(N)\otimes U(1)]/{\bf Z_{N}}$ ones
\be
M^{SU(N)}(n,n)=M^{U(N)}(n,n)~~,~~\forall{n}\in{{\bf Z}_{\geq{0}}}~,
\label{3.9b}
\ee
as it is proven in Appendix B.

\subsection{The Dual form of the $D$-matrix actions.}

Complementary to the reformulation (\ref{3.6}) of the measure, a generic
$D$-matrix action (\ref{2.0}) can be rewritten in a more concise synthetic
form combining $both$ the 'normal' $U_{j}^{i}$-variables $and$ the dual
degrees of freedom. As a result, integrating out $\{U_{\rho}\}$ with the help
of the $S(n)$-formula (\ref{3.6}), the PF $\tilde{X}_{r}$ of (\ref{2.0})
can be expressed in terms of the dual variables only.

It is appropriate to recall first the synthetic representation of the
$SU(N)$ group characters (see e.g. \cite{Moore})
\be
\chi_{R}(U)=\frac{1}{n!}\sum_{\sigma\in S(n)}\chi_{R}(\sigma)
Tr_{n} [D(\sigma)U^{\oplus n}]=
Tr_{n} [D(C_{R})U^{\oplus n}]~,
\label{3.10}
\ee
associated to the set of irreps $R\in{Y^{(N)}_{n}}$ with 
the Young tableaus containing a given number $n$ of boxes.
Eq. (\ref{3.10}) simply rewrites the Frobenius formula \cite{Dr&Zub} in terms
the conventional algebraic notations
\be
Tr_{n} [D(\sigma)U^{\oplus n}]=\sum_{i_{1}i_{2}..i_{n}=1}^{N}
U_{i_{1}}^{i_{\sigma(1)}}U_{i_{2}}^{i_{\sigma(2)}}...
U_{i_{n}}^{i_{\sigma(n)}}=\prod_{k=1}^{n} [tr(U^{k})]^{p_{k}}~,
\label{3.11}
\ee
where $\sigma\in{[1^{p_{1}}2^{p_{2}}...n^{p_{n}}]}$, i.e. $\sigma$ belongs to
the $S(n)$ conjugacy class $[\sigma]$ defined by the associated partition
of $n$: $\sum_{k=1}^{n} k~p_{k}=n$. Note also that the complete
set of $\chi_{R}(U)$ is expressed (see e.g. \cite{Dr&Zub}) with the help of
$(U)^{i}_{j}$-factors, while $(U^{+})^{i}_{j}$ is $not$ engaged.

Topologically, each individual trace
$tr(U^{k})$ in eq. (\ref{3.11}) can be
visualized by the $k$-fold winding of a path around a single base-cycle
associated to the $U$-factor. To generalize the 1-cycle construction (\ref{3.11})
for the case of the $D$-matrix action (\ref{2.0}),
consider first a single 'closed' $q$-loop $tr(U_{\mu_{q}}U_{\nu_{q}}...
U_{\lambda_{q}})$. 
One observes that the latter trace can be visualized now by a path wrapped,
according to the structure of the associated word $w^{(s)}_{q}$,
around the $D$ independent $\rho$-cycles of the base-lattice.
In turn, any ($C(2m)$-cyclic symmetrized) word $\mu\nu...\lambda$
(constrained for simplicity by (\ref{6.20v})) of a length $2m=
2\bigoplus_{\rho=1}^D m_{\rho}$ can be reproduced
\be
  tr(U_{\mu}U_{\nu}... U_{\lambda})=
  {Tr_{2m} [D(\alpha_{\{m\}})
  \bigotimes_{\rho=1}^{D} \left(~(U_{\rho})^{\oplus m_{\rho}}\otimes
(U_{\rho}^{+})^{\oplus m_{\rho}}\right)]}~,
\label{3.17}
\ee
with the help of the equivalence $class$ $[\alpha_{\{m\}}]$ of the $2m$-cycle
permutations $\alpha_{\{m\}}\in{C(2m)}$ defined modulo certain conjugations
(immaterial for our present discussion). More explicitly, the structure of the
r.h.s. of eq. (\ref{3.17}) adopts the pattern (\ref{3.11}) to the presence of
the $2D$ different $U_{\rho}$, $U_{\rho}^{+}$ basis-factors 
\be
  [(U_{1})^{i_{\alpha(1)}}_{j_{1}}...
  (U_{1})^{i_{\alpha(m_{\rho})}}_{j_{m_{\rho}}}
  (U_{1}^{+})^{k_{\alpha(m_{\rho}+1)}}_{l_{m_{\rho}+1}}...
  (U_{1}^{+})^{k_{\alpha(2m_{\rho})}}_{l_{2m_{\rho}}}]\otimes ...
  [...  (U_{D}^{+})^{k_{\alpha(2m)}}_{l_{2m}}],
\label{3.18}
\ee
where the mapping $n\rightarrow{\alpha(n)},~n=1,...,2m,$ defines the
$S(2m)$ permutation $\alpha_{\{m\}}$.

Next, eq. (\ref{3.17}) by the same token
represents a generic $\{w^{(s)}_{k}\}_{\tilde{n}(+)}$
product of traces (with $2m=\tilde{n}(+)$ ) entering the $D$-matrix action
(\ref{2.0}) constrained by
(\ref{6.20v}). For this purpose, one is to choose such
$\sigma({\{w^{(s)}_{k}\}})\in{S(2m)}$ that can be decomposed
into the $ordered$ product $\sigma=P\otimes_{k=1}^{p(\sigma)} c_{n_{k}}$
of the $n_{k}$-cycle permutations $c_{n_{k}}$ reproducing the
trace-product in question. Summarizing, the exponent of the
$D$-matrix action (\ref{2.0}) can be reformulated in the following
synthetic form
\be
e^{-S_{r}(\{U_{\rho}\})}=\sum_{\{n_{\rho}\}} 
Re[\sum_{\sigma}
\psi_{\{n_{\rho}\}}(\sigma) Tr_{4n_{+}} [D(\sigma)
  \bigotimes_{\rho=1}^{D} (U_{\rho})^{\oplus n_{\rho}}\otimes
  (U_{\rho}^{+})^{\oplus n_{\rho}}],
\label{6.0}
\ee
where $n_{\rho}\in{\bf Z_{\geq{0}}}$, $2n_{+}=\sum_{\rho=1}^{D} n_{\rho}$,
and $\sigma\in{S(4n_{+})}$. We remark also that in fact the associated to
(\ref{6.0}) PF $\tilde{X}_{r}$
remains invariant under the substitution $Re[..]\rightarrow{[..]}$, i.e.
one could omit the selection of the real part of the combination in the
rectangular brakets of (\ref{6.0}).

As our attention is restricted to the
solvable deformations of (\ref{1.35}),(\ref{1.36}), we impose extra
condition that $\psi_{\{n_{\rho}\}}(\sigma)$is functionally
parametrized by a set $\{R_{\phi}\in{Y^{(N)}_{n_{\phi}}}\}$ of the relevant
$SU(N)$ irreps
\be
\psi_{\{n_{\rho}\}}(\sigma)=\sum_{\{R_{\phi}\}}e^{-S(\{R_{\phi}\})}~
\psi_{\{n_{\rho}\}}(\sigma|\{R_{\phi}\})~~~,~~~
n_{\rho}=\sum_{\nu\neq{\rho}}^{D-1} n_{\rho\nu}~.
\label{6.0b}
\ee
As a result, the $D$-matrix action (\ref{6.0})/(\ref{6.0b}), being defined in
terms of the $\{\Xi_{\{n({\rho})\}}\}$-set of the $S(4n_{+})$-algebra elements
\be
\Xi_{\{n({\rho})\}}=\sum_{\sigma\in{S(4n(+))}}
\psi_{\{n({\rho})\}}(\sigma)~\sigma
=\sum_{\{R_{\phi}\}} e^{-S(\{R_{\phi}\})}~
\Xi_{\{n({\rho})\}}(\{R_{\phi}\}),
\label{5.1b}
\ee
can be resummed in terms of the alternative set of the (properly normalized)
operators $\Xi_{\{n({\rho})\}}(\{R_{\phi}\})\in{S(4n_{+})},~
\phi\in{\{\mu\nu\},\{\rho\}},$ weighted by a numerical factor
$e^{-S(\{R_{\phi}\})}$. Note also the advantage of choosing the $SU(N)$
option of the $D$-matrix models
(\ref{2.0}), (\ref{6.20v}), where the $SU(N)$ link-variables can be extended
(according to (\ref{3.9b})) to the $U(N)$ ones. In this
way, we combine the simpler structure of the $U(N)$ 1-link integral
(\ref{3.6}) with the more compact pattern (\ref{3.10}) of the $SU(N)$
characters (implicitly entering the action through the operators
(\ref{5.1b})).

Summarizing, in the nonabelian case the dual representation
introduces the $extended$ (compared to (\ref{2.63})) set of the $dual$
variables: the integer-valued $\{\lambda(\phi)\}$-fields parametrizing
the relevant irreps $R_{\phi}$ are complemented by the elements of the
$\bigotimes_{n} S(n)$-algebra. In the particular case of the $SU(N)$ GLR
generating functionals (\ref{1.35}),(\ref{1.36}), the pertinent dual degrees
of freedom fit the pattern
\be
\{~\bigotimes_{n} S(n)~~;~~\{\lambda\}\in{~
[{\bf Z}^{N}/S(N)]^{\oplus \frac{D(D-1)}{2}}\otimes 
[{\bf Z}^{N}/S(N)]^{\oplus D}
~}~\},
\label{3.4}
\ee
where, for a given $\phi\in{\{\mu\nu\},\{\rho\}}$, each
$\{\lambda(\phi)\}$-sector is composed of
the $SU(N)$ sets of $N-1$ $nonnegative$ integers
$\{\lambda^{SU(N)}\}=\{\lambda_{1}>\lambda_{2}>..>\lambda_{N-1}>0\}$.
The latter enter the scene through the relevant Young idempotents
$C_{R_{\phi}}$.

\subsection{$D$-matrix amplitudes v.s. $Tr_{4n(+)}$-characters.}

Let us now put together the dual pattern (\ref{3.6}) of the  measure and
the synthetic representation (\ref{6.0}) of the $D$-matrix action. To take
advantage of
the large $N$ limit, we change the relative order between the integration
$\prod_{\{\rho\}}dU_{\rho}$ and the summation (\ref{6.0b}) over
$\{R_{\phi}\}$: for a $finite$ $N$, one is
to integrate out the $\{U_{\rho}\}$-variables (containing the $O(N^{2})$
degrees of freedom) in the first place. To justify this interchange, the
weight-function $\psi_{\{n_{\rho}\}}(\sigma)$ in eq. (\ref{6.0})
should provide, for any finite $N$, with the $absolute$ and $uniform$ in
$U_{\rho}\in{U(N)}$ convergence of the $\{n_{\rho}\}$-series.
In the dual reformulation (\ref{3.4}),
the integral over $\{U_{\rho}\}$ is traded for the sums (\ref{3.6}).
Therefore, the summation over the $\bigotimes_{n} S(n)$-elements (for each
particular
$\{n_{\rho}\}$) is to be performed $prior$ to the remaining sum over the
$O(N)$ degrees of freedom parametrizing irreps $R_{\phi}$ involved into
(\ref{6.0b}) and (\ref{3.6}). The resulting effective $\{R_{\phi}\}$-theory
(generalizing the functional (\ref{6.3p})) can be approached, at least in
principle, by the subsequent large $N$ saddle-point analysis.

Following the proposed strategy, one is to express the $D$-matrix PF
$\tilde{X}_{r}$ as the weighted sum of the master-integrals 
\be
Tr_{4n(+)}[D(A_{\{n_{\rho}\}})]=\int 
Tr_{4n(+)} [D(\Xi_{\{n_{\rho}\}})D(\{U_{\rho}\otimes U_{\rho}^{+}\})]
~\prod_{\tilde{\rho}=1}^{D}dU_{\tilde{\rho}},
\label{5.1}
\ee
equal to the
character of the corresponding master-elements $A_{\{n(\rho)\}}$ 
(belonging to
the the $S(4n(+))$-algebra) in the $tensor$ representation.  
 In eq. (\ref{5.1}) we have defined (following eq.
(\ref{6.0}))
\be
D(\{U_{\rho}\otimes U_{\rho}^{+})\})\equiv{
\bigotimes_{\rho=1}^{D}\left(~(U_{\rho})^{\oplus n({\rho})}
\otimes (U_{\rho}^{+})^{\oplus n(\rho)}\right)}~,
\label{5.1v}
\ee
where $\Xi_{\{n({\rho})\}}\in{S(4n(+))}$ is introduced in eq. (\ref{5.1b})
so that the block (\ref{5.1})
is associated in
eq. (\ref{6.0}) to the subset of the terms
summed up for
the particular partition $\{n_{\rho}\equiv{n(\rho)}\}$ of a given
$2n_{+}\equiv{2n(+)}$. To
derive the master-element $A_{\{n(\rho)\}}$, in eq. (\ref{5.1})
the result of the $D$ different $U_{\rho}$-integrations (\ref{3.6}) 
is to be represented as an operator embedded to act in the same enveloping
$S(4n(+))$-space where {\it both} $\Xi_{\{n(\rho)\}}$ {\it and} the
complementary block (\ref{5.1v}) (being considered as the operator) act.

For this purpose, let us first specify a $S(4n(+))$-basis
suitable to accomplish our program.
As it is reviewed in Appendix A, each individual $\hat{U}_{\rho}$ or
$\hat{U}^{+}_{\rho}$ acts on the associated elementary
subspace $\hat{U}_{\rho}|i_{-}(\rho)>=(U_{\rho})_{i}^{j}|j_{-}(\rho)>,~
\hat{U}^{+}_{\rho}|i_{+}(\rho)>=(U^{+}_{\rho})_{i}^{j}|j_{+}(\rho)>,~
i,j=1,...,N$. Thus, a given realization of the $S(4n(+))$-basis
is to be constructed as a properly {\it ordered} outer product of the
elementary building blocks $|i_{\pm}(\rho)>$.
In particular, the ordering of the $\{U_{\rho}\}$-factors
in eqs. (\ref{5.1}),(\ref{3.18}) is associated to the following basis
\be
|\tilde{I}_{4n(+)}>=\bigotimes_{\rho=1}^{D}|I_{2n(\rho)}>~~~;~~~
|I_{2n(\rho)}>=|I^{(+)}_{n(\rho)}>\otimes |I^{(-)}_{n(\rho)}>,
\label{5.13}
\ee
\be
|I^{(\pm)}_{n(\rho)}>=\bigotimes_{\nu\neq{\rho}}^{D-1}
|I^{(\pm)}_{n(\rho\nu)}>
~;~
|I^{(\pm)}_{n(\rho\nu)}>=|i_{\pm}(\rho)>^{\oplus n(\rho\nu)}
|i_{\pm}(\nu)>^{\oplus n(\rho\nu)},
\label{5.13b}
\ee
where $2n(+)=\sum_{\rho=1}^{D} n({\rho})$.

Returning to the $S(4n(+))$ representation of the $D$ 1-link integrations,
it is more effective to employ the alternative, $S(2n)$-reformulation of the
$S(n)\otimes S(n)$ formula (\ref{3.6}) (see Appendix B) which
in the $|I_{2n}>=|I^{(+)}_{n}>\otimes |I^{(-)}_{n}>$ basis reads
\be
\int dU D(U)^{j_{1}...j_{n}}_{i_{1}...i_{n}}
D(U^{+})^{j_{n+1}...j_{2n}}_{i_{n+1}...i_{2n}}=
D(\Phi_{2n}\Gamma(2n)(\Lambda^{(-1)}_{n}\otimes{\hat{1}_{[n]}}))
^{\{j^{\oplus 2n}\}}_{\{i^{\oplus 2n}\}},
\label{5.11}
\ee
where $|I^{(\pm)}_{n}>=|i_{\pm}>^{\oplus n}$ matches with
$|I^{(\pm)}_{n(\rho)}>$ of eq. (\ref{5.13b}), and $\hat{1}_{[n]}$ denotes
the 'unity'-permutation of the $S(n)$ group. As for the operator
$\Lambda^{(-1)}_{n}\in{S(n)}$, it is defined by eq. (\ref{3.8}), while
\be
D(\Gamma(2n))^{\{j^{\oplus 2n}\}}_{\{i^{\oplus 2n}\}}=
\sum_{\delta\in{S(n)}}
{D(\delta^{-1})}
^{j_{1}...j_{n}}_{i_{1}...i_{n}}\otimes
{D(\delta)^{j_{n+1}...j_{2n}}_{i_{n+1}...i_{2n}}}\in{S(2n)}~.
\label{5.12}
\ee
Let us restore the $\rho$-labels, i.e. $n\rightarrow{n(\rho)}$.
For a given link $\rho$, the left and the right
$S(n(\rho))$-subblocks of $\Gamma(2n(\rho))$ in eq. (\ref{5.12}) act
respectively on the 'chiral', $|I^{(+)}_{n(\rho)}>$, and the 'antichiral',
$|I^{(-)}_{n(\rho)}>$, $S(n(\rho))$-subspaces of $|I_{2n(\rho)}>$ entering
eq. (\ref{5.13}). The same convention is used for the $S(n(\rho))$-subblocks
in the direct product $(\Lambda^{(-1)}_{n}\otimes{\hat{1}_{[n]}})$
entering
eq. (\ref{5.11}). The remaining $S(2n)$-operator $\Phi_{2n}$,
being considered in the alternative ordered basis for each
$|I_{2n(\rho)}>$-subsector 
\be
|I_{2n(\rho)}>\rightarrow{|\tilde{I}_{2n(\rho)}>}
=(|i_{+}(\rho)>\otimes |i_{-}(\rho)>)^{\oplus n(\rho)}~,
\label{5.14}
\ee
(with $|i_{\pm}(\rho)>^{\oplus {n}(\rho)}=
\otimes_{\nu\neq{\rho}}^{D-1}|i_{\pm}(\rho)>^{\oplus n(\rho\nu)}$), 
takes the simple form of the outer product of the
2-cycle permutations $c_{2}\in{C(2)}$
\be
\Phi_{2n(\rho)}={(c_{2})}^{\oplus n(\rho)}\in{S(2n(\rho))}~~~;~~~
c_{2}:\{12\}\rightarrow{\{21\}}~,
\label{5.15}
\ee
with each $c_{2}\in{S(2)}$ acting on the 'elementary'
sector $|i_{+}(\rho)>\otimes |i_{-}(\rho)>$.  Combining all the pieces
together, the master-element $A_{\{n(\rho)\}}$ is
supposed to be constructed as the composition of the involved (into eqs.
(\ref{5.1b}) and (\ref{5.11})) $S(n(\phi))$-valued operators $embedded$ to
act in the $common$ 'enveloping' $S(4n(+))$-space.  

Let us apply this algorithm in order to rederive the GLR pattern (\ref{6.3p})
directly evaluating the master-integrals (\ref{5.1}) associated to the 
eigenvalue-models (\ref{1.35}) or (\ref{1.36}). Given the basis (\ref{5.13}),
the corresponding to (\ref{1.36}) operator $\Xi_{\{n(\rho)\}}$
 assumes (after identification
$S^{(1)}(\{R_{\phi}\})=S(\{R_{\phi}\})$ in eq. (\ref{5.1b}))
the following form
\be
\Xi^{(1)}_{\{n(\rho)\}}(\{R_{\phi}\})=\bigotimes_{\rho=1}^{D}
K^{(1)}_{2n(\rho)}~~;~~
K^{(1)}_{2n(\rho)}=(C_{R_{\rho}}\bigotimes{[\otimes_{\nu\neq{\rho}}^
{D-1}C_{R_{\rho\nu}}]}),
\label{5.4}
\ee
while the substitution $C_{R_{\rho}}\rightarrow
{[\otimes_{\nu\neq{\rho}}^{D-1}C_{R_{\rho\nu}}]}$
reproduces $K^{(2)}_{2n(\rho)}$ of the option (\ref{1.35}). 
By definition, each $C_{R_{\rho\nu}}$-factor
in eq. (\ref{5.4}) is postulated to act on
the corresponding  $|I^{(-)}_{n(\rho\nu)}>$ subspace of
(\ref{5.13b}), while each $C_{R_{\rho}}$-factor acts on the associated
$|I^{(+)}_{n(\rho)}>$ subspace.

Actually, a preliminary variant of the GLR pattern (\ref{6.3p}) can be
deduced from (\ref{5.4}) already at this step. For this purpose, one is to
combine the
the peculiar structure of the dual $\Xi^{(m)}_{\{n(\rho)\}}(\{R_{\phi}\})$-
operator (\ref{5.4}) with the invariance
of any $D$-matrix action (\ref{6.0}) with respect to the substitution
of $\Xi_{\{n(\rho)\}}$ by its 'twisted' partner
\be
\Xi_{\{n(\rho)\}}\rightarrow{\sum_{\{\sigma_{\pm}(\rho)\}}
[\sigma_{+}\otimes \sigma_{-}]^{-1}\Xi_{\{n({\rho})\}}~
[\sigma_{+}\otimes \sigma_{-}]}~~,~~
\sigma_{\pm}=\otimes_{\rho=1}^{D} \sigma_{\pm}(\rho),
\label{6.2z}
\ee
where $\sigma_{\pm}(\rho)\in{S(n(\rho))}$ is postulated to act on the
corresponding $|I^{(\pm)}_{n(\rho)}>$-subspace of the alternative $S(4n(+))$
basis
\be
|I'_{4n(+)}>=|I^{(+)}_{2n(+)}>\bigotimes |I^{(-)}_{2n(+)}>~~;~~
|I^{(\pm)}_{2n(+)}>=\otimes_{\rho=1}^{D}|I^{(\pm)}_{n(\rho)}>~,
\label{6.8}
\ee
with $|I^{(\pm)}_{n(\rho)}>$ being defined by eq. (\ref{5.13b}).
Performing the substitution (\ref{6.2z}) for
$\Xi^{(m)}_{\{n(\rho)\}}(\{R_{\phi}\})$ defined by (\ref{5.4}), we finally
employ the fusion rules of the Young idempotents (see Appendix D)
\be
\sum_{\delta\in{S(n_{+})}}\frac{
[\delta~ (\otimes_{k=1}^{p} C_{R_{k}})~ \delta^{-1}]}{(n_{+})!}=
\bigoplus_{R_{+}\in{Y_{n_{+}}}}
L^{(p)}_{R_{+}|\{R_{k}\}} ~C_{R_{+}}~~;~~n_{+}=
\sum_{k=1}^{p} n_{k},
\label{6.2}
\ee
to arrive at the structure which foreshadows the $D$-products (\ref{6.3p}) of
the GLR coefficients.

As for the invariance (\ref{6.2z}), it is a consequence of the basic
commutativity
$[D(\sigma),U^{\oplus n}]=0,~\forall{\sigma \in{S(n)}}$ (see Appendix A). The
latter ensures that the ordered product (\ref{5.1v})
(associated to the $|\tilde{I}_{4n(+)}>$-basis (\ref{5.13})) is equal
to its '$twisted$' counterpart
\be
\bigotimes_{\rho=1}^{D} \sum_{\{\sigma_{\pm}(\rho)\}}
(\sigma_{+}(\rho)\otimes \sigma_{-}(\rho))^{-1}[(U_{\rho})^{\oplus n_{\rho}}
\otimes (U_{\rho}^{+})^{\oplus n_{\rho}}]
(\sigma_{+}(\rho)\otimes \sigma_{-}(\rho)),
\label{6.2c}
\ee
where $\sigma_{\pm}(\rho)\in{S(n(\rho))}$. Being rewritten in the
$|{I'}_{4n(+)}>$ basis (\ref{6.8}), it matches with
(\ref{6.2z}). Remark also that, compared to eq. (\ref{6.3}), the
decomposition (\ref{6.2}) involves a larger set of the GLR coefficients
of $pth$ order: the involved irreps $R_{\psi}$ are parametrized by the
$S(n_{\psi})$, rather than $SU(N)$, Young tableaus $Y_{n_{\psi}}$ (see
Appendices A and D for the relevant details). 

Returning to the derivation of the master-element $A^{(m)}_{\{n(\rho)\}}$
associated to the eigenvalue-models (\ref{1.35}),(\ref{1.36}), one is to
put together eqs. (\ref{5.1}), (\ref{5.11}) and (\ref{5.4}). Prior to
the
twisting (\ref{6.2z}), it results in
\be
A^{(1)}_{\{n(\rho)\}}=
\otimes_{\rho=1}^{D} \Phi_{2n(\rho)}~\frac{\Gamma(2n(\rho))}{n!}~
\frac{d_{R_{\rho}}~K^{(1)}_{2n(\rho)}}{dimR_{\rho}}~,
\label{5.17}
\ee
while the substitution $K^{(1)}_{2n(\rho)}\rightarrow{
([\otimes_{\nu\neq{\rho}}^{D-1}C_{R_{\rho\nu}}]P_{R_{\rho}}
\bigotimes{[\otimes_{\mu\neq{\rho}}^{D-1}C_{R_{\rho\mu}}]})}$
reproduces $A^{(2)}_{\{n(\rho)\}}$ corresponding to the option (\ref{1.35}).
Let us show how, inside the $Tr_{4n(+)}$ character (\ref{5.1}), the
expression (\ref{5.17})
can be further simplified to end up with the GLR functional (\ref{6.3p}).
First, according to the invariance (\ref{6.2z}),
the operator $\Xi^{(m)}_{\{n({\rho})\}}(\{R_{\phi}\})$ 
 can be substituted 
by its $(\sigma_{+}\otimes \sigma_{-})$-twisted partner. 
Combining it the identity which in the $|I_{2n}>$-basis reads
\be
\Gamma(2n)~[\sigma_{+}\otimes \sigma_{-}]=
[\sigma_{-}\otimes \hat{1}_{[n]}]~\Gamma(2n)~
[\sigma_{+}\otimes \hat{1}_{[n]}]~~,~~\forall{\sigma_{\pm}\in{S(n)}},
\label{6.2v}
\ee
inside the character one can substitute
${A}^{(1)}_{\{n(\rho)\}}\rightarrow{\tilde{A}^{(1)}_{\{n(\rho)\}}}$ where 
\be
\tilde{A}^{(1)}_{\{n(\rho)\}}=\otimes_{\rho=1}^{D} \Phi_{2n(\rho)}
\frac{\Gamma(2n(\rho))}{n(\rho)!} \sum_{R_{\rho}\in{Y_{n(\rho)}^{(N)}}}
\left(\frac{L^{(D-1)}_{R_{\rho}|\{R_{\rho\nu}\}}~C_{R_{\rho}}}
{dimR_{\rho}}\otimes
\hat{1}_{[n(\rho)]}\right).
\label{6.2b}
\ee 

Let us note that the invariance with respect to the latter substitution
independently follows from $[P_{R},\sigma]=0,~\forall{\sigma},$ and the
identity (\ref{B.24}) of Appendix B. The latter formulas
ensure that in eq. (\ref{5.17}) the product of the operators, complementary to
$\Xi^{(1)}_{\{n(\rho)\}}(\{R_{\phi}\})$, belongs to the $center$ of
$[\otimes_{\rho=1}^{D}S(n(\rho))]\bigotimes [\otimes_{\rho=1}^{D}S(n(\rho))]$
(with each of the $S(n(\rho))$-factors
acting on the corresponding $|I^{(\pm)}_{n(\rho)}>$-subspaces of (\ref{6.8})),
i.e. the subgroup corresponding to the conjugations (\ref{6.2z}). In turn,
together with (\ref{6.2v}) it justifies the validity of the twisting
(\ref{6.2b}) of ${A}^{(1)}_{\{n(\rho)\}}$.
Similarly, for the option $A^{(2)}_{\{n(\rho)\}}$ one
reproduces the product of the GLR coefficients which matches (after some
auxiliary trick discussed in Section 5.1)
with the $m=2$ case of (\ref{6.3p}).

To ensure that the remaining factors in (\ref{6.2b})
conspire to reproduce exactly the correct GLR functional (\ref{6.3p}), one is
 to use first the defining property of $\Phi_{2n}$ which in the
basis $|I_{2n}>=|I^{(+)}_{n}>\otimes |I^{(-)}_{n}>$ assumes the form 
\be
Tr_{2n}[D((\sigma_{+}\otimes \sigma_{-})~\Phi_{2n})]=
Tr_{n}[D(\sigma_{+}\sigma_{-})]~~,~~\forall{\sigma_{\pm}}\in{S(n)}~,
\label{6.2y}
\ee
where $\sigma_{\pm}$ acts on the corresponding $|I^{(\pm)}_{n}>$-subspace.
Combining (\ref{6.2y}) with the completeness condition
\be
\frac{1}{n!}\sum_{\sigma\in{S(n)}}
\chi_{R_{1}}(\sigma \rho)\chi_{R_{2}}(\sigma^{-1} \alpha)=
\delta_{R_{1},R_{2}}\frac{\chi_{R_{1}}(\rho {\alpha})}{d_{R_{1}}}
\label{6.2vv}
\ee
and the projection-formula, which tells that $Tr_{n} [D(P_{R}\sigma)]$
(where $P_{R}=d_{R}C_{R}$ and ${R}\in{Y_{n}^{(N)}}$) is nonzero only
for ${R}\in{Y_{n}}$ when
\be
Tr_{n} [D(P_{R}\sigma)]=dimR~\chi_{R}(\sigma)~~,~~\forall{R}\in{Y_{n}^{(N)}}~,
\label{6.2vvxx}
\ee
we finally rederive (\ref{6.3p}).

In conclusion, let us compare how the nonabelian fusion-rules' constraints
are realized in the original and dual representations. 
In the former, 'normal' formulation (\ref{6.3p}), the
conditions on the admissible irreps $R_{\phi}$ are imposed by the
products of the GLR coefficients. On the side of the dual
representation, the fusion rules are realized
in terms of the $\bigotimes_{n} S(n)$ degrees of
freedom (\ref{3.4}) combined into the $Tr_{4n_{+}}$-characters (\ref{5.1}).
Owing to the inherent combinations of the Young
idempotents $C_{R_{\phi}}$, algebraically the symmetric groups' variables
play the role of the $S(n)$-valued $Lagrange$ $multipliers$ (absent in the
abelian case (\ref{2.61})).

\section{GLR-computable $D$-matrix PFs.}
\setcounter{equation}{0}

Now we are in a position to formulate the concept of the symmetry, which
underlies the resolution of the fusion rule algebra in
terms of the GLR coefficients (\ref{6.3}) for a subvariety of the
$D$-matrix
systems (\ref{2.0}),(\ref{6.20v}) graded by the rank $k=1,...,D$ of the
conjugation-invariance (\ref{2.0d}). Since this symmetry is $not$ manifest in
the original formulation (\ref{2.0}) (based on the word-parametrization of
the traces involved), one can view it as the $hidden$ symmetry of the system.

The idea is to induce the $k<D$ GLR solvable systems generalizing the
$k=D$ operator $\Xi^{(m)}_{\{n_{\rho}\}}(\{R_{\phi}\})$ (defined
for $m=1$ in eq. (\ref{5.4})) in such a way so that the following basic
property of the latter operator is preserved. Namely, the
invariance of a generic $D$-matrix action under the twisting
(\ref{6.2z}) of $\Xi_{\{n_{\rho}\}}$ must be transformed into that under the
complementary twisting (akin to (\ref{6.2})) of the
$\{C_{R_{\phi}}\}$-factors entering
$\Xi^{(m)}_{\{n_{\rho}\}}(\{R_{\phi}\})$
in the master-integral (\ref{5.1}). One easily observes that the required
$symmetry$ (with respect to the 'switching' of the
$(\sigma_{+}\otimes \sigma_{-})$-twist
(\ref{6.2c}) from the $\{U_{\rho^{\pm}}\}$- to
the $\{C_{R_{\phi}}\}$-block) holds true provided in eq. (\ref{5.1})
$\Xi_{\{n_{\rho}\}}$ is chosen in the following form
\be
\Xi_{\{n_{\rho}\}}=\sum_{\{R_{\phi}\}} 
e^{-E(\{R_{\phi}\})}
~\Xi^{(m)}_{\{n_{\rho}\}}(\{R_{\phi}\})~
\Psi_{\{n_{\rho}\}}(\{R_{\phi}\}),
\label{6.9}
\ee
where $\Xi^{(m)}_{\{n_{\rho}\}}(\{R_{\phi}\})$ is determined by eq. (5.4),
$E(\{R_{\phi}\})$ is some numerical weight-factor, while
$\phi\in{\{\mu\nu\},\{\rho\}}$
and $m=1,2$. As for $\Psi_{\{n_{\rho}\}}$, it can be an $arbitrary$ element
belonging to the $center$ of $any$ subalgebra $\tilde{S}\bigotimes \tilde{S}$
of $S(2n_{+})\bigotimes S(2n_{+})$ (in what follows we employ the
$|I'_{4n_{+}}>$-basis (\ref{6.8}))
which contains the subsubalgebra
\be
[\otimes_{\rho=1}^{D} S(n_{\rho})]~
\bigotimes ~[\otimes_{\rho=1}^{D} S(n_{\rho})]
\label{6.9v}
\ee
inherent in the twisting (\ref{6.2c}). More generally, $\Psi_{\{n_{\rho}\}}$
should commute with
$any$ element of the group-product (\ref{6.9v}) that can be concisely
formalized by the pattern
\be
\Psi_{\{n_{\rho}\}}=\frac{1}{[n(\tilde{S})]^{2}}
\sum_{\{\sigma_{\pm}\in{\tilde{S}}\}}
[\sigma_{+}\otimes \sigma_{-}]^{-1}
M_{\{n_{\rho}\}}(\{R_{\phi}\})~[\sigma_{+}\otimes \sigma_{-}]~,
\label{6.1}
\ee
where $n(\tilde{S})$ denotes the number of the elements in $\tilde{S}$, while
$M_{\{n_{\rho}\}}$ {\it a priori} may be a generic element of the
$S(4n_{+})$ algebra (consistent with the convergence of the final
summation (\ref{6.0}) over $\{n_{\rho}\}\in{[{\bf Z_{\geq{0}}}]^{D}}$).

Depending on the choice of $M_{\{n_{\rho}\}}$ and the admissible subgroup
$\tilde{S}\in{S(2n_{+})}$, the resulting (via (\ref{6.9}) and (\ref{5.1}))
$D$-matrix action (\ref{2.0}) is endowed with the
conjugation-symmetry (\ref{2.0d}) of a different rank $k$. In what follows,
we concentrate on the simplest case when $\tilde{S}={S(2n_{+})},~
n(\tilde{S})=(2n_{+})!,$ resulting in the $k=1$ subvariety of the 
models (\ref{2.0}) which can be viewed as the {\it non}eigenvalue deformations
of the two basic $k=D$ systems (\ref{1.35}),(\ref{1.36}). To develope an
intuition in what
is going on, let us obtain the explicit form of the associated $D$-matrix
actions with the
arguments restricted to the
subspace of the coinciding link-variables: $U_{\rho}=U,~
\forall{\rho}=1,...,D$.
In this case, there appears the $larger$ symmetry under the switching of
the $extended$ $(S(2n_{+})\otimes S(2n_{+}))$-twisting (which generalizes
(\ref{6.2c})) 
\be
\sum_{\{\sigma_{\pm}\in{S(2n_{+})}\}}
(\sigma_{+}\otimes \sigma_{-})^{-1}[(U)^{\oplus 2n_{+}}
\otimes (U^{+})^{\oplus 2n_{+}}]~
(\sigma_{+}\otimes \sigma_{-}),
\label{6.2cc}
\ee
onto the complementary $\{C_{R_{\phi}}\}$-block of (\ref{6.9}).
To be more specific, consider the $m=2$ option of (\ref{6.9}) and
choose the
simplest separable form of $M_{\{n_{\rho}\}}$ 
\be
M_{\{n_{\rho}\}}=
M^{(1)}_{2n_{+}}(\{R_{\phi}\})\otimes
M^{(2)}_{2n_{+}}(\{R_{\phi}\})~~,~~
M^{(1,2)}_{2n_{+}}\in{S(2n_{+})},
\label{6.1b}
\ee
where $\phi\in{\{\mu\nu\},\{\rho\}}$. Taking into account the identity
\be
\sum_{\sigma} \frac{\sigma
(\bigotimes_{\rho=1}^{D} \{P_{R_{\rho}}
[\otimes_{\nu\neq{\rho}}^{D-1} C_{R_{\rho\nu}}]\})\sigma^{-1}}
{(2n_{+})!}=
\sum_{R_{+}} (\prod_{\mu=1}^{D} L^{(D-1)}_{R_{\mu}|\{R_{\mu\nu}\}})
L^{(D)}_{R_{+}|\{R_{\rho}\}}C_{R_{+}},
\label{6.18}
\ee
(where $\sigma\in{S(2n_{+})},~R_{+}\in{Y_{2n_{+}}}$) and summing up
$\sum_{R_{\rho}\in{Y_{n_{\rho}}}} P_{R_{\rho}}=1$, one
obtains
\be
e^{-\tilde{S}^{(2)}_{r}(\{U\})}=\sum_{\{R^{(q)}_{\phi}\}}
e^{-E(\{R_{\mu\nu}\})}
H(\{R^{(q)}_{\phi}\}) \prod_{p=1}^{2}
\frac{\chi_{R^{(p)}_{+}}(M^{(p)}_{2n_{+}})\chi_{R^{(p)}_{+}}(U^{(p)})}
{d_{R^{(p)}_{+}}},
\label{6.1c}
\ee
\be
H(\{R^{(q)}_{\phi}\})=\prod_{p=1}^{2}
[~L^{(D)}_{R^{(p)}_{+}|\{R^{(p)}_{\rho}\}}
(\otimes_{\mu=1}^{D}~L^{(D-1)}_{R^{(p)}_{\mu}|\{R_{\mu\nu}\}})~]~,
\label{6.1d}
\ee
where $U^{(2)}\equiv{U^{+}},~U^{(1)}\equiv{U}$ and
$\phi\in{+,\{\rho\},\{\mu\nu\}}$.

Next, the analysis of the continuum limit (in the associated induced gauge
theory) will require the knowledge of the $explicit$ relation between the two
weights: $E(\{R_{\phi}\})$ (entering
$e^{-\tilde{S}_{r}(\{U\})}
\equiv{e^{-\tilde{S}_{r}(\{U_{\rho}\})}|_{\{U_{\rho}=U\}}}$)
and $S(\{R_{\phi}\})$ involved into the $GLR$ computable $D$-matrix
partition function (\ref{6.3p}). To derive a transparent example of
such a relation, we concentrate on the deformations of (\ref{1.36}) and
specify the operator $M_{\{n_{\rho}\}}$ further in the form
generalizing (\ref{6.1b})
\be
M_{\{n_{\rho}\}}(\{R_{\phi}\})=
[(\tilde{M}^{(1)}_{2n_{+}}F_{2n_{+}})\otimes \tilde{M}^{(2)}_{2n_{+}}]~
\frac{\Gamma(4n_{+})}{(2n_{+})!},
\label{6.30}
\ee
where $\Gamma(4n_{+})\in{S(4n_{+})}$ is defined by eq.
(\ref{5.12}) (with each $S(2n_{+})$-subblock acting on the corresponding 
$|I^{(\pm)}_{2n_{+}}>$-subspace of (\ref{6.8})). The auxiliary factor
\be
F_{2n_{+}}=\sum_{R_{+}\in{Y_{2n_{+}}^{(N)}}} 
e^{-E(\{R_{\mu\nu}\},R_{+})+E(\{R_{\mu\nu}\})}
~P_{R_{+}}~.
\label{6.10}
\ee
is introduced to trade $E(\{R_{\mu\nu}\})$ in the final
amplitudes for its $R_{+}$ {\it dependent} counterpart (and for simplicity we
consider $\{R_{\rho}\}$-independent weights).

To begin with, employing (\ref{6.18}) and the identities listed in the
end of the previous section, one easily obtains for the associated to
(\ref{6.30}) deformation of the eigenvalue-action (\ref{1.36}) (considered
for the coinciding arguments)
\be
e^{-\tilde{S}^{(1)}_{r}(\{U\})}=\sum_{\{R^{(q)}_{\phi}\}}
e^{-E}~Q(\{R^{(q)}_{\phi}\})~
\frac{\chi_{R_{+}}(\tilde{M}^{(1)}_{2n_{+}}\tilde{M}^{(2)}_{2n_{+}})}
{d^{3}_{R_{+}}}|\chi_{R_{+}}(U)|^{2},
\label{6.31}
\ee
\be
Q(\{R^{(q)}_{\phi}\})=
\prod_{p=1}^{2} L^{(D)}_{R_{+}|\{R^{(p)}_{\rho}\}}~
\left(\otimes_{\mu=1}^{D}~L^{(D-1)}_{R^{(2)}_{\mu}|\{R_{\mu\nu}\}}\right)~.
\label{6.32}
\ee

Let us now turn to the evaluation of
the PF $\tilde{X}_{r}$ corresponding to (\ref{6.31}). Similarly to eq.
(\ref{6.2b}), the master-integral (\ref{5.1}) associated to (\ref{6.30}) is
expressed in terms of the following $S(4n_{+})$-algebra element
$\tilde{A}_{\{n_{\rho}\}}$
\be
\sum_{\{R_{\phi}\}} e^{-E}
[\tilde{M}^{(1)}_{2n_{+}}\otimes \tilde{M}^{(2)}_{2n_{+}}]
\Phi_{4n_{+}} \{\sum_{\{\sigma_{\pm}\}}
[\sigma_{+}\otimes \sigma_{-}]\frac{W_{4n_{+}}}{([2n_{+}]!)^{2}}
[\sigma_{+}\otimes \sigma_{-}]^{-1}\},
\label{6.33}
\ee
\be
W_{4n_{+}}=[\left(P_{R_{+}}\otimes_{\rho=1}^{D}C_{R_{\rho}}
\frac{L^{(D-1)}_{R_{\rho}|\{R_{\rho\nu}\}}}{dimR_{\rho}}\right)
\bigotimes{\hat{1}_{[n_{\rho}]}}]\frac{\Gamma(4n_{+})}{[2n_{+}]!}
[\otimes_{\rho=1}^{D}\frac{\Gamma(2n_{\rho})}{[2n_{\rho}]!}],
\label{6.34}
\ee
where the second sum runs over
$\sigma_{\pm}\in{S(2n_{+})}$ and
$\Phi_{4n_{+}}=\otimes_{\rho=1}^{D}\Phi_{2n_{\rho}}$. To simplify eq.
(\ref{6.33}) further, we first note that the factor $\otimes_{\rho=1}^{D}
{\Gamma(2n_{\rho})}/{[2n_{\rho}]!}$ in the
$(\sigma_{+}\otimes \sigma_{-})$-twisted operator $W_{4n_{+}}$ can be 
substituted by the $\hat{1}_{[4n_{+}]}$-unity employing the proper change of
the variables. Indeed, let $\Gamma(4n_{+})$ and $\Gamma(2n_{\rho})$ are
defined by eq. (\ref{5.12}) in terms of $\delta\in{S(2n_{+})}$ and
$\delta_{\rho}\in{S(n_{\rho})}$ elements respectively. Making (for a given
$\delta_{\rho}$) the two shifts
$\delta\rightarrow{\delta[\otimes_{\rho=1}^{D}\delta^{-1}_{\rho}}],~
\sigma_{-}\rightarrow{\sigma_{-}[\otimes_{\rho=1}^{D}\delta^{-1}_{\rho}}]$,
one eliminates the dependence of (\ref{6.33}) on
$\{\delta_{\rho}\}$.

Second, the resulting form of eq. (\ref{6.33}) can be shown to be 
invariant under the
substitution of $[P_{R_{+}}\otimes_{\rho=1}^{D}C_{R_{\rho}}]$ by
\be
{\sum_{\tilde{\sigma}\in{S(2n_{+})}} \tilde{\sigma}
\frac{[P_{R_{+}}\otimes_{\rho=1}^{D}C_{R_{\rho}}]}{[2n_{+}]!}~
\tilde{\sigma}^{-1}}=L^{(D)}_{R_{+}|\{R_{\rho}\}}~C_{R_{+}}
\label{6.35}
\ee
made in the 'chiral' $S(2n_{+})$-block of the combination in the first
rectangular brakets of (\ref{6.34}). To this aim, one is to perform
(for a given $\tilde{\sigma}$) the following composition of the shifts and the
conjugation: $\sigma_{+}\rightarrow{\sigma_{+}\tilde{\sigma}^{-1}},~
\delta\rightarrow{\tilde{\sigma}\delta \tilde{\sigma}^{-1}},~
\sigma_{-}\rightarrow{\sigma_{-}\tilde{\sigma}^{-1}}$.
Altogether, the tensor-dependent part
of eq. (\ref{6.33}) can be rewritten as
\be
(C_{R_{+}}\tilde{M}^{(1)}_{2n_{+}}\otimes
\tilde{M}^{(2)}_{2n_{+}})~
\Phi_{4n_{+}} \sum_{\{\sigma_{\pm}\}}
(\sigma_{+}\otimes \sigma_{-})~\frac{\Gamma(4n_{+})}{([2n_{+}]!)^{3}}~
(\sigma_{+}\otimes \sigma_{-})^{-1},
\label{6.36}
\ee
where $\sigma_{\pm}\in{S(2n_{+})}$.
Employing (\ref{6.2vvxx}) together with (\ref{6.2y}) and
introducing
$\tilde{\sigma}_{+}=[\sigma^{-1}_{-}\tilde{M}^{(2)}_{2n_{+}}]\sigma_{+}$
instead of $\sigma_{+}$, one transforms eq. (\ref{6.36}) into
\be
\frac{dimR_{+}}{d_{R_{+}}}
\sum_{\delta,\tilde{\sigma}_{+},\sigma_{-}}
\frac{\chi_{R_{+}}(\delta^{-1}\tilde{\sigma}_{+}\delta\tilde{\sigma}^{-1}_{+}
\{\sigma^{-1}_{-} \tilde{M}^{(2)}_{2n_{+}}\tilde{M}^{(1)}_{2n_{+}}
\sigma_{-}\})}{([2n_{+}]!)^{3}},
\label{6.38}
\ee
where the sums over $\delta,\tilde{\sigma}_{+},\sigma_{-}$ run over
the ${S(2n_{+})}$-group elements.
Finally, combining $\sum_{\{\delta,\sigma\in{S(n)}\}}
\chi_{R}(\delta\sigma\delta^{-1}\sigma^{-1})/(n!)^{2}=1/d_{R}$ (see
\cite{Gr&Tayl}) together with
\be
\sum_{\sigma\in{S(n)}} \frac{\sigma B_{n} \sigma^{-1}}{n!}=
\sum_{R\in{Y_{n}}}\frac{\chi_{R}(B_{n})}{d_{R}}~P_{R}~~,~~
\forall{B_{n}}\in{S(n)},
\label{6.38v}
\ee
(see Appendix D), and with
$\chi_{R_{1}}(P_{R_{2}}\sigma)=\delta_{R_{1},R_{2}}\chi_{R_{1}}(\sigma)$, we
finally obtain
\be
\tilde{X}_{r}=\sum_{\{R_{\phi}\}} e^{-E(\{R_{\mu\nu}\},R_{+})}~
\tilde{Q}(\{R_{\phi}\})~
\frac{\chi_{R_{+}}(\tilde{M}^{(1)}_{2n_{+}}\tilde{M}^{(2)}_{2n_{+}})~
dimR_{+}}
{d^{3}_{R_{+}}\otimes_{\rho=1}^{D}~dimR_{\rho}},
\label{6.39}
\ee
\be
\tilde{Q}(\{R_{\phi}\})=L^{(D)}_{R_{+}|\{R_{\rho}\}}~\left(~
\otimes_{\mu=1}^{D} L^{(D-1)}_{R_{\mu}|\{R_{\mu\nu}\}}~\right)~,
\label{6.39b}
\ee
where the sum runs over the $SU(N)$ irreps $\{R_{\phi}\},~
\phi\in{+,\{\rho\},\{\mu\nu\}}$.

In conclusion, one observes that both in eq. (\ref{6.39}) and in eq.
(\ref{6.31}) there appears
the same factor $K(R_{+})=
{\chi_{R_{+}}(\tilde{M}^{(1)}_{2n_{+}}\tilde{M}^{(2)}_{2n_{+}})}/
{d^{3}_{R_{+}}}$
$violating$ the invariance under the ${\bf Z_{2}}$-conjugation:
$\otimes_{\phi} R_{\phi}\leftrightarrow{
\otimes_{\phi} \bar{R}_{\phi}}$, where $\phi\in{\{\mu\nu\},\{\rho\},+}$.
To retain this auxiliary symmetry (and make contact with
$S(\{R_{\phi}\})$ in (\ref{4.50}) of Section 5), we redefine
\be
e^{-E(\{R_{\mu\nu}\},R_{+})}K(R_{+})=e^{-\tilde{E}(\{R_{\mu\nu}\},R_{+})}=
e^{-S(\{R_{\phi}\})}
\frac{\otimes_{\rho=1}^{D}dimR_{\rho}}{dimR_{+}},
\label{6.39c}
\ee
postulating that $\tilde{E}(\{R_{\mu\nu}\},R_{+})$ is ${\bf Z_{2}}$-invariant.

\section{Mapping onto the induced gauge theory.}
\setcounter{equation}{0}

In \cite{Dub1} we have developed the algorithm that associates to the
$k=D$ eigenvalue models like (\ref{1.35}),(\ref{1.36}) the $D$-dimensional
induced gauge theory in such a way that the correspondence (\ref{2.10})
(between the the PFs) holds true. Our present purpose is to induce,
preserving (\ref{2.10}), gauge theories from the generic $k\leq{D}$
$D$-matrix systems (\ref{2.0}) (invariant under (\ref{2.0d}) and
(\ref{2.9})) including those belonging to the GLR computable variety defined
via eqs. (\ref{6.9}),(\ref{6.1}).

To begin with, taking (\ref{1.35}) as an example, let us briefly review the
algorithm designed in \cite{Dub1} for the $k=D$ $D$-matrix models. It
consists of the two steps. First, one
employs the large $N$ saddle-point method to prove that the $SU(N)$ system
(\ref{1.35}) is reduced (eliminating the space-time dependence) from the
following $D$-dimensional eigenvalue-system. The latter is defined associating
to each site ${\bf x}$ (of $L^{D}$ lattice) the factor
\be
\sum_{\{R_{\mu\nu}\}} e^{-S}
\prod_{\{\mu\nu\}} \chi_{R_{\mu\nu}}({U}_{\mu}({\bf x}))
\chi_{R_{\mu\nu}}({U}_{\nu}({\bf x+\mu}))
\chi_{R_{\mu\nu}}({U}^{+}_{\mu}({\bf x+\nu}))
\chi_{R_{\mu\nu}}({U}^{+}_{\nu}({\bf x})),
\label{6.25c}
\ee
where $S\equiv{S(\{R_{\mu\nu}\})}$ and
$R_{\mu\nu}\equiv{R_{\mu\nu}({\bf x}))}$. Observe that the correspondence
between
(\ref{1.35}) and (\ref{6.25c}) implies the particular choice of the mapping
$\{U_{\rho}\}\rightarrow{\{U_{\rho}({\bf z})\}}$ between the link-variables
entering the PFs $\tilde{X}_{r}$ and $\tilde{X}_{L^{D}}$ respectively.

The constructed in this way intermediate system (\ref{6.25c}) is invariant
under the local $[U(N)]^{\oplus D}$ conjugation-symmetry
\be
U_{\rho}({\bf z})\rightarrow
{h^{+}_{\rho}({\bf z})~U_{\rho}({\bf z})~h_{\rho}({\bf z})}~~,~~
h_{\rho}\in{U(N)}~~,~~\rho=1,...,D,
\label{2.50}
\ee
combined with the reduced gauge symmetry with respect to the center $T$ of
the Lie group
\be
U_{\rho}({\bf z})\rightarrow
{H^{+}({\bf z})~U_{\rho}({\bf z})~H({\bf z+\rho})}~~,~~
H({\bf z})\in{T}~~,~~T={\bf Z_{N}}~,
\label{2.50b}
\ee
complemented by the global $[{\bf Z_{N}}]^{\oplus D}$-invariance
\be
[T]^{\oplus D}:~U_{\rho}({\bf z})\rightarrow{t_{\rho}U_{\rho}({\bf z})}~~,~~
t_{\rho}\in{T={\bf Z_{N}}}~.
\label{2.9x}
\ee
The latter two symmetries substantiate consistency of the second step: the
gauge theory is induced from the system (\ref{6.25c}) through the
'gauge transformation'
\be
U_{\rho}({\bf z})\rightarrow{
\tilde{G}^{+}({\bf z})~U_{\rho}({\bf z})~\tilde{G}({\bf z+\rho})}~~,~~
\rho=1,...,D,
\label{2.0z}
\ee
introducing the auxiliary $SU(N)$ scalar field $\tilde{G}({\bf z})$ assigned
to the lattice sites. Integration over $\tilde{G}({\bf z})$ with the Haar
measure (normalized by $\int d\tilde{G}({\bf z})=1$) results \cite{Dub1} in
the associated effective theory with the manifestly gauge-invariant action
$\tilde{S}_{eff}(\{U_{\rho}({\bf z})\})$.

The pairing between the local conjugation-symmetry (\ref{2.50}) and
its global $k=D$ counterpart (\ref{2.0d}) is crucial for maintaining
the large $N$ correspondence (\ref{2.10}) between (\ref{6.25c}) and its
reduced partner (\ref{1.35}). Being intended to induce a gauge theory
from the $k<D$ {\it non}eigenvalue $D$-matrix models
(\ref{2.0})/(\ref{6.9}), we propose to map preliminary these models
onto the associated effective $D$-matrix eigenvalue-theories. Then, to
maintain the correspondence (\ref{2.10}) with the latter effective theory,
an appropriate modification of the associated pattern (\ref{6.25c}) of the
intermediate $D$-dimensional eigenvalue-theory will be found.
Finally, the mapping (\ref{2.0z}) will produce, as
previously, an induced gauge theory.

To fulfil this program, let us start with the construction of the effective
eigenvalue-theory associated to a given $k<D$ model (\ref{2.0}) which
for simplicity is restricted to satisfy (\ref{6.20v})
with $n_{\rho}$ being additionally constrained by 
\be
n_{\rho}=\sum_{\nu\neq{\rho}}^{D-1} n_{\rho\nu}~~,~~
n_{\mu\nu}\in{\bf Z_{\geq{0}}}~.
\label{6.20vvx}
\ee
For this purpose, in (\ref{2.0}) one is to rewrite
$U_{\rho}=\Omega_{\rho}~
diag[e^{i\omega(\rho)}]~\Omega_{\rho}^{+}$ and then integrate over
$\Omega_{\rho},~\rho=1,...,D,$ employing (\ref{3.9b}) and the the
decomposition of the $U(N)$ measure
\be
\int_{U(N)} dU=\int d\Omega~\prod_{k=1}^{N}\int_{-\pi}^{+\pi}
\frac{d\omega_{k}}{2\pi}~
\prod_{i<j} |2sin(\frac{\omega_{i}-\omega_{j}}{2})|^{2}~,
\label{1.2d}
\ee
The point is that the
integrations over the $right$-cosets
$\Omega_{\rho}\in{U(N)/[U(1)]^{\oplus N}}$ can be extended,
introducing the auxiliary matrix $\tilde{T}_{\rho}\in{[U(1)]^{\oplus N}}$,
to those over $W_{\alpha}$ spanning the {\it full} $U(N)$ group-manifold
(so that the dual representation (\ref{3.6}) of the 1-link integral is
applicable to this preliminary mapping).

Second, the remaining integrals over
${T}_{\rho}\equiv{diag[e^{i\omega(\rho)}]}
\in{[U(1)]^{\oplus N}}$ by the same token can be promoted (introducing
$\Upsilon_{\rho}\in{U(N)/[U(1)]^{\oplus N}}$)
to those over $\tilde{U}_{\rho}\in{U(N)}$ which altogether reads
$d\Omega_{\rho}\rightarrow{d(\Omega_{\rho}\tilde{T}_{\rho})}
\equiv{dW_{\rho}},~
dT_{\rho}\rightarrow{d(\Upsilon_{\rho}T_{\rho}\Upsilon^{+}_{\rho})}
\equiv{d\tilde{U}_{\rho}}$.
In terms of this extended set of the variables, the PF of a generic $D$-matrix
model (\ref{2.0}) can be rewritten
\be
\tilde{X}_{r}=\int_{U(N)} e^{-S_{eff}(\{\tilde{U}_{\rho}\})}
\prod_{\tilde{\rho}=1}^{D} d\tilde{U}_{\tilde{\rho}}=
\int_{SU(N)} e^{-S_{eff}(\{\tilde{U}_{\rho}\})}
\prod_{\tilde{\rho}=1}^{D} d\tilde{U}_{\tilde{\rho}}
\label{1.30}
\ee
as the PF of the associated effective eigenvalue-theory manifestly invariant
under the $k=D$ conjugation-symmetry (\ref{2.0d})
\be
S_{eff}(\{\tilde{U}_{\rho}\})=
-ln[\int_{U(N)}~e^{-\tilde{S}_{r}(\{W_{\rho}\tilde{U}_{\rho}W^{+}_{\rho}\})}
\prod_{\tilde{\rho}=1}^{D} dW_{\tilde{\rho}}~],
\label{1.30b}
\ee
where the particular normalization $\int_{U(N)} dU=1$ of the Haar measure is
used. In the derivation of the second, $SU(N)$ form of $\tilde{X}_{r}$ in
(\ref{1.30}), we employ the invariance of $S_{eff}(\{\tilde{U}_{\rho}\})$
under (\ref{2.9}) (i.e. (\ref{6.20v})) that allows to apply the identity
(\ref{3.9b}).
To return from (\ref{1.30}) to the original representation, one
absorbs $\Upsilon_{\rho}$ by the opposite 'shift'
$W_{\rho}\rightarrow{W_{\rho}\Upsilon^{+}_{\rho}}$ of the
$W_{\rho}$-variables and then employs the commutativity
$\tilde{T}_{\rho}~diag[e^{i\omega(\rho)}]~\tilde{T}^{+}_{\rho}=
diag[e^{i\omega(\rho)}]$. As for the maximal $k=D$ symmetry (\ref{2.0d}) of
(\ref{1.30b}), it follows from
the possibility to reabsorb the $\Psi_{\rho}$-rotations in the same way as we
have done for $\Upsilon_{\rho}$.

The action of the resulting effective $SU(N)$
eigenvalue-theory can be defined by in the following form
\be
e^{-\tilde{S}_{r}(\{\tilde{U}_{\rho}\})}=
\sum_{\{n_{\rho}\}}
\sum_{\{R^{(q)}_{\rho}\in{Y^{(N)}_{n_{\rho}}}\}}
e^{-A(\{R^{(q)}_{\rho}\})}Re[~
\prod_{\rho=1}^{D} \chi_{R^{(-)}_{\rho}}(\tilde{U}_{\rho})
\chi_{R^{(+)}_{\rho}}(\tilde{U}^{+}_{\rho})~],
\label{2.0vv}
\ee
where $A\equiv{A(\{R^{(q)}_{\rho}\})}$ and the associated to
$\{R^{(q)}_{\rho}\}$ numbers of boxes
$\{n_{\rho}(q)\}$ satisfy (owing to (\ref{6.20v})) $n_{\rho}(1)=n_{\rho}(2)
\equiv{n_{\rho}},~\forall{\rho}$.

To reconstruct an associated intermediate $D$-dimensional eigenvalue-system,
let us first compare the pattern (\ref{2.0vv}) with the one
(where $U_{\rho}^{(+)}\equiv{U_{\rho}^{+}},~
U_{\rho}^{(-)}\equiv{U_{\rho}}$) 
\be
\sum_{\{n_{\mu\nu}\}}
\sum_{\{R^{(q)}_{\mu\nu}\in{Y^{(N)}_{n_{\mu\nu}}}\}}
e^{-A(\{R^{(q)}_{\mu\nu}\})}Re[
\prod_{\mu\nu=1}^{\frac{D(D-1)}{2}} \prod_{q=\pm}
\chi_{R^{(q)}_{\mu\nu}}(U^{(q)}_{\mu})
\chi_{R^{(q)}_{\mu\nu}}(U^{(q)}_{\nu})],
\label{2.0vvv}
\ee
that generalizes (\ref{1.35})
remaining compatible with the algorithm employed in \cite{Dub1} to
induce a gauge theory. The $D$-dimensional eigenvalue-system corresponding to
(\ref{2.0vvv}) can be deduced from (\ref{6.25c}) substituting each          
$\mu\nu$-block of the characters by the more general block (where
$R^{(q)}_{\mu\nu}=R^{(q)}_{\mu\nu}({\bf x})$)
\be
\chi_{R^{(1)}_{\mu\nu}}({U}_{\mu}({\bf x}))
\chi_{R^{(1)}_{\mu\nu}}({U}_{\nu}({\bf x+\mu}))
\chi_{R^{(2)}_{\mu\nu}}({U}^{+}_{\mu}({\bf x+\nu}))
\chi_{R^{(2)}_{\mu\nu}}({U}^{+}_{\nu}({\bf x})),
\label{6.25cc}
\ee                                     
with the overall weight being given by
$S\equiv{S(\{R^{(q)}_{\mu\nu}({\bf x})\})}$.

One observes that for $D\geq{3}$ the pattern (\ref{2.0vv}) of the effective
eigenvalue-theory, according to Frobenius formula (\ref{3.10}),(\ref{3.11}),
generically can $not$ be reproduced in terms of (\ref{2.0vvv}) (resulting
from (\ref{6.25cc}) after the large $N$ SP reduction of the space-time
dependence of ${U}_{\mu}({\bf z}),R^{(q)}_{\mu\nu}({\bf x})$).
To adjust the algorithm of \cite{Dub1} to the more general $D\geq{3}$ family
(\ref{2.0vv}), we note first that the latter systems can be reduced
(eliminating the space-time dependence) from the following $D$-dimensional
eigenvalue-systems. The latter are defined associating to each site ${\bf x}$
(instead of (\ref{6.25c})) the factor
\be
\sum_{\{n_{\rho}\}}
\sum_{\{R^{(q)}_{\rho}\in{Y^{(N)}_{n_{\rho}}}\}}
e^{-A(\{R^{(q)}_{\rho}\})} ~Re[~
\prod_{\rho=1}^{D} \prod_{q=\pm}
\chi_{R^{(q)}_{\rho}}({U}^{(q)}_{\rho}({\bf x}))~],
\label{6.25ccc}
\ee
which provides with the mapping $\{U_{\rho}\}\rightarrow
{\{U_{\rho}({\bf z})\}}$
alternative to the one encoded in the pairing between (\ref{2.0vvv}) and
(\ref{6.25cc}). By the same token as in \cite{Dub1}, the PF $\tilde{X}_{L^{D}}$
of (\ref{6.25ccc}) is related to that $\tilde{X}_{r}$ of (\ref{2.0vv})
through the large $N$ correspondence (\ref{2.10}).

Next, application of the second mapping (\ref{2.0z}) converts the
$D$ dimensional eigenvalue-system (\ref{6.25ccc}) into an induced gauge
theory. Altogether, this prescription provides with the algorithm which
induces a gauge model from a generic $D$-matrix system
(\ref{2.0})/(\ref{6.20v}) including the $k\leq{D}$ family
(\ref{6.9}),(\ref{6.1}) with the GLR computable PF.
The subtlety is that the intermediate eigenvalue-system (\ref{6.25cc})
in addition is invariant under the (finite $N$) {\it local}
$[{\bf Z_{N}}]^{\oplus D}$ symmetry
\be
[{\bf Z_{N}}]^{\oplus D}:~U_{\rho}({\bf z})\rightarrow
{t_{\rho}({\bf z})U_{\rho}({\bf z})}~~,~~
t_{\rho}({\bf z})\in{{\bf Z_{N}}}~,
\label{2.50m}
\ee
'much larger' than the ${\bf Z_{N}}$ gauge invariance (\ref{2.50b}).
In turn, symmetry (\ref{2.50m}) is present in the induced via
(\ref{2.0z}) gauge theory that is known to set zero the average of $any$
Wilson loop $ W_{C}(U)=tr(U_{\mu}({\bf x})U_{\nu}({\bf x+\mu})...
U_{\rho}({\bf x-\rho}))$ provided the corresponding to the contour $C$
(minimal) area does {\it not} vanish. Therefore, it calls for a modification
of the prescription to get rid of the unwanted invariance (\ref{2.50m})
keeping (\ref{2.50b}) intact.

To circumvent this problem, we propose the following synthetic algorithm
defining the mapping $\{U_{\rho}\}\rightarrow{\{U_{\rho}({\bf z})\}}$ for
the link-variables of (\ref{2.0vv}).
First, one is to use the Frobenius formula (\ref{3.10}),(\ref{3.11})
expanding the characters in (\ref{2.0vv}) in terms of the trace products. In
the resulting sum, consider a particular term (substituting
$\tilde{U}_{\rho}\rightarrow{U_{\rho}}$)
\be
\sim{\prod_{\rho=1}^{D}~\prod_{k}tr((U_{\rho})^{k})]^{p^{-}_{k}(\rho)}
~\prod_{\tilde{k}}tr((U^{+}_{\rho})^{\tilde{k}})]^{p^{+}_{\tilde{k}}(\rho)}}
~~,~~\sum_{k} kp^{\pm}_{k}(\rho)=n_{\rho}~,
\label{3.10b}
\ee
containing total amount $n_{\rho}$ of the $U_{\rho}$ (or, equally,
$U^{+}_{\rho}$) factors. Second, let us separate (in a way specified below)
the product (\ref{3.10b}) into two blocks splitting the partititions
$[1^{p^{\pm}_{1}(\rho)}2^{p^{\pm}_{2}(\rho)}...]$ according to
$p^{\pm}_{k}(\rho)=
l^{\pm}_{k}(\rho)+f^{\pm}_{k}(\rho)$ so that
\be
\sum_{k} kl^{\pm}_{k}(\rho)=m^{(1)}_{\rho}~~,~~
\sum_{k} kf^{\pm}_{k}(\rho)=m^{(2)}_{\rho}~~,~~
m^{(1)}_{\rho}+m^{(2)}_{\rho}=n_{\rho}~.
\label{3.10c}
\ee

Given (\ref{3.10c}), perform (with the help of the second Frobenius formula (\ref{A.12})) multiple
Fourier expansion of the first, $\{l^{\pm}_{k}(\rho)\}$-block
\be
\prod_{\{\rho,k,q\}}tr((U^{(q)}_{\rho})^{k})]^{l^{q}_{k}(\rho)}=
\sum_{\{R^{(q)}_{\rho}\}} e^{-B_{1}(\{R^{(q)}_{\rho}\})}
\prod_{\{\rho,q\}}\chi_{R^{(q)}_{\rho}}({U}^{(q)}_{\rho}).
\label{3.10d}
\ee
Let the link-variables $U_{\rho}$ in
(\ref{3.10d}) be mapped onto ${U_{\rho}({\bf z}})$ of the intermediate
$D$-dimensional eigenvalue-system in compliance with the pattern (\ref{6.25ccc}).
We postulate that the set $\{m^{(1)}_{\rho}\}$, minimizing the 
function $\sum_{\rho=1}^{D} [m^{(1)}_{\rho}]^{2}$, is constrained
by the following condition. There should exist a set
$\{m_{\mu\nu}\in{\bf Z_{\geq{0}}}\}$ of $D(D-1)/2$
integers so that the second $\{f^{\pm}_{k}(\rho)\}$ block can be represented
in the form
\be
\prod_{\{\rho,k,q\}} 
tr((U^{(q)}_{\rho})^{k})]^{f^{q}_{k}(\rho)}=
\sum_{\{R^{(q)}_{\mu\nu}\}}
e^{-B_{2}} \prod_{\{\mu\nu,q\}}
\chi_{R^{(q)}_{\mu\nu}}(U^{(q)}_{\mu})
\chi_{R^{(q)}_{\mu\nu}}(U^{(q)}_{\nu}),
\label{3.10e}
\ee
(where $B_{2}\equiv{B_{2}(\{R^{(q)}_{\mu\nu}\})}$) that matches with the
pattern (\ref{2.0vvv}). In eq. (\ref{3.10e}), it is supposed that
$R^{(q)}_{\mu\nu}\in{Y^{(N)}_{m_{\mu\nu}}}$ with
$m^{(1)}_{\rho}=\sum_{\rho\neq{\nu}}^{D-1} m_{\rho\nu}$. As a result, the
latter condition evidently allows to map $\{U_{\rho}\}$ in (\ref{3.10e}) onto
$\{U_{\rho}({\bf z})\}$ according to the pattern (\ref{6.25cc}).

Upon a reflection, the above prescription remains essentially ambiguous. To
fix the freedom of choosing the $\{m_{\mu\nu}\in{\bf Z_{\geq{0}}}\}$ set,
we impose extra constraint that the latter integers minimize
$\sum_{\{\mu\nu\}}[m_{\mu\nu}-n_{\mu\nu}]^{2}$ (where $n_{\mu\nu}$ enters
(\ref{6.20vvx})). In case if there remains (accidental) residual
ambiguity, one is to symmetrize over all the admissible options. Summarizing,
we have formulated the algorithm to map a given $D$-matrix model (\ref{2.0})
(constrained by (\ref{6.20v}),(\ref{6.20vvx})) onto the intermediate
$D$-dimensional eigenvalue-system which is invariant under
(\ref{2.50b}),(\ref{2.50}).
Applying to the latter the final mapping (\ref{2.0z}), we induce
the theory invariant under the conventional $SU(N)$ gauge invariance.

\section{Continuum limit of the induced theories.}
\setcounter{equation}{0}

The analysis of the continuum limit in the lattice gauge theories, induced
from the GLR computable $k<D$ models (\ref{2.0})/(\ref{6.20vvx}), essentially
follows the route employed in \cite{Dub1} for gauge theories induced
directly from the $k=D$models like (\ref{1.35}).
In what follows we briefly sketch the major steps with the emphasis on
a few novel details.

The idea is to take advantage of the fact that the infinite correlation
length in a gauge theory is supposed to
entail the following effect. Namely, the link-variables
$U_{\rho}({\bf z})$ are supposed to be {\it localized} (modulo (\ref{2.9x})
and the gauge transformations) in the infinitesimal vicinity, scaling as
$O(N^{(0)})$, of the group-unity element $\hat{1}$. The gauge-invariant
representation of this condition implies the existence of some $O(N^{0})$
functional $\tilde{g}^{2}N\equiv
{\tilde{g}^{2}(\{g_{k}\})}N\rightarrow{0}$ (of the relevant coupling
constants $\{g_{k}\}$) so that
\be
\lim_{N\rightarrow{\infty}}
\lim_{\tilde{g}^{2}N\rightarrow{0}} 
|\frac{1}{N}<tr[U(pl)]>-1|\sim{O(\tilde{g}^{2}N)}~,
\label{7.2xx}
\ee
where $U(pl)={U}_{\mu}({\bf x}){U}_{\nu}({\bf x+\mu})
{U}^{+}_{\mu}({\bf x+\nu}){U}^{+}_{\nu}({\bf x})$ stands for the holonomy
around an elementary plaquette in an arbitrary $\mu\nu$-plane.
Let us fix the 'maximal tree' gauge \cite{Dr&Zub} putting
$U_{\rho}({\bf z})=\hat{1}$ on
a largest possible $tree$ (made of the links) which by definition does not
contain nontrivial 1-cycles.
Then, introducing the quantum fluctuations $A^{ab}_{\rho}({\bf z})
=-iln[U^{ab}_{\rho}({\bf z})]$, the required localization can be formulated
in the large $N$ limit in the form
\be
\lim_{N\rightarrow{\infty}} \lim_{\tilde{g}^{2}N\rightarrow{0}}
<[A^{ab}_{\rho}({\bf z})]^{2}>\sim
{O(\tilde{g}^{2})}~\bmod (\ref{2.9x})~,~\forall{a,b=1,...,N}.
\label{7.2x}
\ee

Following \cite{Dub1}, we intend to prove that in the induced theory
the constraint (\ref{7.2x}) is fulfilled if in the associated $D$-matrix
model (defined for definiteness by eqs. (\ref{6.30}),(\ref{6.39c}))
the condition 
\be
\lim_{N\rightarrow{\infty}}
\lim_{\tilde{g}^{2}N\rightarrow{0}}
<[A^{ab}_{\rho}]^{2}>\sim{
O(\tilde{g}^{2})}~~\bmod (\ref{2.9})~,
\label{7.2b}
\ee
(where $A^{ab}_{\rho}=-iln[U^{ab}_{\rho}]$), is valid for any given
$a,b=1,...,N$. In turn, to ensure scaling
(\ref{7.2b}) we first represent the $D$-matrix PF 
$\lim_{N\rightarrow{\infty}} \tilde{X}^{(m)}_{r}$ as the $(mD)$th power
of the effective 1-matrix $SU(N)$ theory formulated in terms of irreps $R$
\be
\lim_{N\rightarrow{\infty}} \tilde{X}^{(m)}_{r}=\lim_{N\rightarrow{\infty}}
[~{\sum_{R}}''~e^{-S^{(m)}(R|D)}~]^{mD}~.
\label{1.2b}
\ee
which is valid \cite{Dub1} provided $-ln[\tilde{X}^{(m)}_{r}] \sim{O(N^{2})}$
and the weight $S^{(m)}(\{R_{\phi}\})$ in eq. (\ref{6.3p}) (defining
$\tilde{X}^{(m)}_{r}$) is invariant under the group-product
\be
S(D)\otimes S(D(D-1)/2)\otimes {\bf Z_{2}}~~;~~{\bf Z_{2}}:~ 
\otimes_{\phi} R_{\phi}\leftrightarrow{
\otimes_{\phi} \bar{R}_{\phi}},
\label{6.3pc}
\ee
combining the separate permutations within the two sets ($\{\mu\nu\}$
and $\{\rho\}$) of the irrep-indices $\phi\in{\{\rho\},\{\mu\nu\}}$
together with the $simultaneous$ conjugation of $all$ the involved into
eq. (\ref{6.3p}) irreps $R_{\phi}$.

As we will demonstrate, the condition (\ref{7.2b}) can be reformulated as the
following constraint on the saddle-point (SP) values of the
$\lambda_{j}\in{\bf Z}$ fields canonically parametrizing the irreps $R$ in
the effective 1-matrix system (\ref{1.2b}).
Namely, the SP values $\lambda^{(0)}_{i}=N\bar{\lambda}^{(0)}_{i}$
should approach 'infinity' according to the
complementary $scaling$-condition
\be
\lim_{N\rightarrow{\infty}}
\lim_{\tilde{g}^{2}N\rightarrow{0}} |\lambda^{(0)}_{i}|\sim
{O(N/[\tilde{g}N^{\frac{1}{2}}])}~~\Longleftrightarrow{~~
|\bar{\lambda}^{(0)}_{i}|\sim{O([\tilde{g}N^{\frac{1}{2}}]^{-1})}},
\label{7.1}
\ee
with the functional $\tilde{g}(\{g_{k}\})$ (which enters
(\ref{7.2xx}),(\ref{7.2b})) tending to zero.

\subsection{The effective $N\rightarrow{\infty}$ 1-matrix theory.}

To prove that (\ref{7.2b}) yields (\ref{7.2xx}), we first
derive the 1-matrix representation
(\ref{1.2b}) of the large $N$ PF (\ref{6.3p})
associated to the $D$-matrix model (\ref{2.0}) specified by
(\ref{6.30}),(\ref{6.39c}). As we will see, the sum ${\sum_{R}}''$ in
(\ref{1.2b}) is in fact constrained \cite{Dub1} by the condition 
that both $n(R)$ and $n(\bar{R})$ must be nonnegative multiples of $(D-1)$
(where $n(R)$ is the number of boxes in the Young tableau $Y^{(N)}_{n(R)}$
associated to $R$).

Upon a reflection, the pattern (\ref{6.39b}) suggests to
start with a little bit more specific $m=1$ form of (\ref{6.3p})
\be
\tilde{X}_{r}=\sum_{\{R_{\phi}\}}
e^{-S(\{R_{\phi}\})}~L^{(D)}_{R_{+}|\{R_{\rho}\}}~
\left(~\otimes_{\mu=1}^{D}~L^{(D-1)}_{R_{\mu}|\{R_{\mu\nu}\}}~\right)~,
\label{4.50}
\ee
where each sum over $R_{\phi}\in{Y_{n_{\phi}}^{(N)}},~\phi\in{\{\mu\nu\},~
\{\rho\},~+},$ runs over the $\phi$-species of the $SU(N)$
irreps. As for the weight-factor $S(\{R_{\phi}\})$,
it is defined in (\ref{6.39c}) being invariant under the group-product
(\ref{6.3pc}), where ${\bf Z_{2}}$ symmetry is extended for 
$\phi=+,\{\rho\},\{\mu\nu\}$. Next, 'integrating out'
in eq. (\ref{4.50}) the auxiliary $R_{+}$ variable, one brings it into the
required $m=1$ form (\ref{6.3p}) with the identification
\be
e^{-S(\{R_{\rho}\},\{R_{\mu\nu}\})}=\sum_{R_{+}}
e^{-S(R_{+},\{R_{\rho}\},\{R_{\mu\nu}\})}~L^{(D)}_{R_{+}|\{R_{\rho}\}}~.
\label{3.40}
\ee

Returning to the reduction of (\ref{6.3p}) to (\ref{1.2b}), it is built on
the localization of
the large $N$ summations over $\{R_{\rho}\}\otimes\{R_{\mu\nu}\}$ on the
solution $\{R^{(0)}_{\rho}\}\otimes\{R^{(0)}_{\mu\nu}\}$ of
the corresponding saddle-point equations. We refer to \cite{Dub1}
for the discussion of these equations, and now simply assert the 
properties of the solution in the case when the constraints (\ref{6.3pc}) are
additionally imposed. To be more specific, we select the option when
the effective 1-matrix system in eq. (\ref{1.2b}) is reduced
to the simplest $solvable$ class of the $SU(N)$ or $U(N)$ models with
$S(R|D)$ being defined as
\be
e^{-S(\{\lambda\})}=|dimR(\{\lambda\})|^{q}~
e^{-\sum\limits_{n=1}^{M_{0}}\sum\limits_{\tilde{r}\in{Y_{2n}}}
g_{\tilde{r}(\{p\})}
\prod\limits_{k=1}^{2n}
[\sum\limits_{i=1}^{N}(\lambda_{i}-\frac{N-1}{2})^{k}]^{p_{k}}}~,
\label{4.1}
\ee
where $q>0$, and in the $SU(N)$ case the set $\{g_{\tilde{r}(\{p\})}\}$ is
supposed to maintain invariance of $S(\{\lambda\})$ under the translations
$\lambda_{i}\rightarrow {\lambda_{i}+m}$. The latter is to match with the
fact that $U(N)$ irreps are labelled by
a set of $N$ integers $\lambda^{U(N)}_{i}$ (constrained by
$\sum_{i=1}^{N-1} (\lambda_{i}-(N-1)/2)=\lambda_{N} \bmod N$)
\be
\{\lambda^{U(N)}\}=\{\lambda_{1}+\lambda_{N}>...>
\lambda_{N-1}+\lambda_{N}>\lambda_{N}\}\in{[{\bf Z}^{\oplus N}/S(N)]}
\label{2.3}
\ee
generated from the $SU(N)$-set of $N-1$ $nonnegative$ integers
$\{\lambda^{SU(N)}\}=\{\lambda_{1}>\lambda_{2}>..>\lambda_{N-1}>0\}$
by the extra integer number $\lambda_{N}\geq{0}$ or $\lambda_{N}<0$.
As for the sum $\sum_{\tilde{r}}$ in the exponent of (\ref{4.1}), it runs
over the irreps $\tilde{r}\equiv{\tilde{r}(\{p\})}\in{Y_{2n}}$ of the even
symmetric group $S(2n)$
(labelled by the partititions $\{p\}=[1^{p_{1}}2^{p_{2}}...2n^{p_{2n}}]$ of
$2n$: $\sum_{k=1}^{2n} kp_{k}=2n$) with $n\leq{M_{0}}\in{{\bf Z_{\geq{1}}}}$.
Under these conditions, the
saddle-point $SU(N)$-set $\{R^{(0)}_{\rho}\}\otimes\{R^{(0)}_{\mu\nu}\}$,
\be
R^{(0)}_{\rho}=R^{(0)}=\bar{R}^{(0)}~,~\forall{\rho}~~~;~~~
R^{(0)}_{\mu\nu}=R_{2}^{(0)}=\bar{R}^{(0)}_{2}~,~\forall{\mu\nu},
\label{3.43b}
\ee
is supposed to be $unique$, $\{\rho\}\otimes\{\mu\nu\}$ independent
respectively, and {\it selfdual}. Consequently, the generating functional
(\ref{6.3p}) is equivalent in the large $N$ limit to the $reduced$ system
resulting after the identification
\be
R_{\mu\nu}\equiv{R_{2}}\in{Y_{n_{2}}^{(N)}}~,~\forall{\mu\nu}~~~;~~~
R_{\rho}\equiv{R}\in{Y_{n}^{(N)}}~,~\forall{\rho}~,
\label{3.43}
\ee
with the remaining summations over $R,~R_{2}$ being localized on the same
saddle-point values (\ref{3.43b}).

The effective action for $R,R_{2}$, resulting from the reduction
(\ref{3.43}), contains (according to (\ref{6.3p})) the $Dth$ power of
$L^{(D-1)}_{R|R^{\oplus D-1}_{2}}$. To simplify this
expression further, one can employ the ($\gamma=D$ case of the) identity
\cite{Dub1}
\be
\lim_{N\rightarrow{\infty}}
\int \prod_{\alpha=1}^{p} \sum_{R_{\alpha}}~
e^{-S_{p}(\{R_{\beta}\})}=
\lim_{N\rightarrow{\infty}}
[~\int \prod_{\alpha=1}^{p} \sum_{R_{\alpha}}~
e^{-S_{p}(\{R_{\beta}\})/\gamma}~]^{\gamma}~,
\label{3.45}
\ee
valid provided that $\gamma>0$, the saddle-point values of $both$
$e^{-S_{p}(\{R^{(0)}_{\beta}\})}$ $and$
$e^{-S_{p}(\{R^{(0)}_{\beta}\})/\gamma}$ are unique and
$positive$, while the corresponding free energy is $\sim{O(N^{2})}$.
Summarizing, it defines the following $one$-matrix representation of the
large $N$ family (\ref{6.3p})
\be
\lim_{N\rightarrow{\infty}} \tilde{X}_{r}=\lim_{N\rightarrow{\infty}}
[~\int dU~\sum_{R,R_{2}} e^{-\frac{S^{(D)}(R,R_{2})}{mD}}~\chi_{R}(U^{+})~
[\chi_{R_{2}}(U)]^{D-1}~]^{mD},
\label{3.44}
\ee
where the weight $S^{(D)}(R,R_{2})$ is deduced from
$S(\{R_{\rho}\},\{R_{\mu\nu}\})$ of (\ref{6.3p}) through 'dimensional
reduction' (\ref{3.43}). Next, owing to the invariance of (\ref{6.3}) (in
(\ref{6.3p})) under (\ref{2.9}), the $SU(N)$
partition function $\tilde{X}_{r}$ is invariant under the substitution
(\ref{3.9b}) of
the $SU(N)$ link-variables by the $U(N)=[SU(N)\otimes U(1)]/{\bf Z_{N}}$ ones.
Therefore, the sum in (\ref{3.44}) over $SU(N)$ irreps $R$
is effectively constrained by the
${\bf Z_{2}}$-invariant pair of the $U(N)$ conditions (nontrivial in $D>2$)
\be
L^{(D-1)}_{R|R^{\oplus D-1}_{2}}\neq{0}~\Rightarrow{~n(R)=n(R_{2})(D-1)}
~,~~n(\bar{R})=n(\bar{R}_{2})(D-1).
\label{3.46c}
\ee
Here the integers $n(R_{\phi}), n(\bar{R}_{\phi})\in{\bf Z_{\geq{0}}}$ denote
the number of boxes in the ${\bf Z_{2}}$-invariant pair of the Young tableaus
corresponding to (\ref{3.43}):
\be
n(R(\{\lambda\}))=\sum_{i=1}^{N} n_{i}=
\sum_{i=1}^{N}(\lambda_{i}-N+i)~,
\label{3.46m}
\ee
while ${\bf Z_{2}}$ conjugation $R\leftrightarrow{\bar{R}}$ reads:
$\{\lambda_{i}\}\leftrightarrow{\{-\lambda_{N-i+1}+\beta\}}$,
where $\beta^{U(N)}=(N-1)$ and $\beta^{SU(N)}=\lambda_{1}$ in the $U(N)$
and $SU(N)$ cases respectively.

Finally, in order to recast eq. (\ref{3.44}) into the form of eq.
(\ref{1.2b}), let us first introduce 
\be
e^{-H(R|D)}=\sum_{R_{2}} e^{-\frac{S^{(D)}(R,R_{2})}{mD}}~
L^{(D-1)}_{R|R^{\oplus D-1}_{2}}~.
\label{3.48}
\ee
By the same token as in \cite{Dub1}, 
the ${\bf Z_{2}}$-invariant pair of the $D>2$ $SU(N)$ conditions
(\ref{3.46c}) is the only constraint
defining the whole $e^{-H(R|D)}$-family induced from the
$e^{-S^{(D)}(R,R_{2})/mD}$-variety via (\ref{3.48}).
It suggests to factorize the latter constraints out
\be
e^{-H(R|D)}=e^{-S(R|D)}~\sum_{k,\bar{k}\in{\bf Z}}
\delta_{n(R),[D-1]k}~\delta_{n(\bar{R}),[D-1]\bar{k}}~,
\label{3.49z}
\ee
so that, for a $fixed$ $D$, in (\ref{1.2b}) $any$ residual $R$-valued
function $e^{-S(R|D)}$ (consistent with the scaling
$-ln[\tilde{X}_{r}]\sim{O(N^{2})}$ and with ${\bf Z_{2}}$-invariance
$S(R|D)=S(\bar{R}|D)$) can be
induced through (\ref{3.48}) provided the judicious adjustment of
$e^{-S^{(D)}(R,R_{2})/mD}$. The two $periodic$ $Kronecker$ delta-functions are
supposed to be defined \cite{Dub1} via certain '$\varepsilon$-regularization'
of the Poisson resummation formula (with explicit form of the latter being
immaterial for the present discussion).
We note also that similar analysis of the $U(N)$ $GLR$
functionals (\ref{1.35}),(\ref{1.36}) (with the sum running over the $U(N)$
irreps) results in the $U(N)$ counterpart of eq. (\ref{1.2b}).

In conclusion, let us remark that in the selected model (\ref{4.1}) the
appropriate
scaling $-ln[\tilde{X}_{r}]\sim{O(N^{2})},~|\lambda_{j}|\sim{O(N)},$  is
adjusted \cite{Dub1} provided for each $r(\{p\})\in{Y_{2n}}$ the following
pattern is valid
\be
\bar{\lambda}_{j}=\lambda_{j}/N~~,~~
g_{r(\{p\})}=b_{r(\{p\})}N^{\gamma_{r}}~~,~~\gamma_{r(\{p\})}=
2-2n-\sum_{k=1}^{2n} p_{k}~,
\label{4.1b}
\ee
where it is postulated that $b_{r(\{p\})}\sim{O(N^{0})}$.

\subsection{The $\{(<tr[U(pl)]>/N)\rightarrow{\hat{1}}\}$ localization.}

Let us prove that, in the gauge theories induced
via (\ref{2.0z}) from the $k=1$ system (\ref{2.0}) (specified by
(\ref{6.30}),(\ref{6.39c})), the required localization (\ref{7.2xx}) of
$U_{\rho}({\bf z})$ is predetermined by the scaling-condition (\ref{7.1}) for
the $\{\lambda\}$ fields entering (\ref{1.2b}),(\ref{4.1}).

To begin with, we recall that the constraint (\ref{7.1}) is dynamically
fulfilled \cite{Dub1} provided in (\ref{4.1}) all coupling constants (rescaled
according to (\ref{4.1b})) tend to zero: $\{b_{r(\{p\})}\rightarrow{0}\}$.
In the simplest case of the $U(N)$ action (\ref{4.1}) with $M_{0}=1$, the
functional $\tilde{g}^{2}$ is to be introduced by
\be
\sum_{R(\{\lambda\})} |dimR(\{\lambda\})|^{q}~
exp[{-g_{\tilde{r}_{0}}\sum_{i=1}^{N}(\lambda_{i}-\frac{N-1}{2})^{2}}]~~,~~
\tilde{g}^{2}/2=g_{\tilde{r}_{0}}~,
\label{4.1c}
\ee
where $g_{\tilde{r}_{0}}\sim{O(1/N)}$ and $\tilde{r}_{0}=[2^{1}]$.
In a general case (\ref{4.1}),(\ref{4.1b}), one is to
choose $b_{2k}=\limsup{[|b_{r}|]}$ as the largest
$|b_{r(\{p\})}|$ in each $k$-subset of $r\in{Y_{2k}}$. Then,
$N\tilde{g}^{2}$ is equated with the
$\limsup{[(b_{2k})^{\frac{1}{k}}]}$ found among all $k\leq{M_{0}}$
(provided the associated leading terms, by themselves, ensure the
convergence of the $\{\lambda\}$-series in (\ref{4.1})).

Next, building on the results of \cite{Dub1}, one might expect that
the scaling (\ref{7.1}) results in the complementary localization
(\ref{7.2b}) of the link-variables $\{U_{\rho}\}$ in the $D$-matrix systems
(\ref{2.0}) specified by (\ref{6.30}),(\ref{6.39c}). This assertion, in
particular, employs  that the action (of the WC perturbation theory) in the
latter system evidently contains (owing to (\ref{6.31})) the
quadratic in $A^{ab}_{\rho}$ term. In turn, the patterns of
the involved mappings (\ref{1.30}) and (\ref{2.0z}) suggest that (\ref{7.2b})
indeed entails the required localization (\ref{7.2xx}) in the gauge theory
induced from (\ref{2.0})/(\ref{6.30}),(\ref{6.39c}).

To substantiate these statements by an explicit computation, we
consider the large $N$ WC asymptotics
$\tilde{g}^{2}N\rightarrow{0}$ of the properly
normalized partition function (PF)
\be
X^{(in)}_{L^{D}}=\int \prod_{\{\rho,{\bf z}\}}dU_{\rho}({\bf z})
exp[{-S(\{U_{\rho}({\bf z})\})-S(\{\hat{1}\})}]
\label{7.3m}
\ee
associated to a (induced) gauge theory on a cubic lattice with $L^{D}$ sites.
In \cite{Dub1} it is shown that the localization (\ref{7.2xx})
generically results in the power-like large $N$ WC asymptotics
\be
\lim_{N\rightarrow{\infty}}
\lim_{\tilde{g}^{2}N\rightarrow{0}}
X^{(in)}_{L^{D}}(\tilde{g})=[C~\tilde{g}N^{\frac{1}{2}}]
^{(D-1) N^{2}L^{D}}~~,~~C>0~,
\label{7.4b}
\ee
where the $\int dU=1$ normalization of the Haar measure
is used, and $C$ is a model dependent constant. Thus, our aim is to prove
that (\ref{7.4b}) is valid in the specific case of the gauge theories induced
from the $D$-matrix models (\ref{2.0})/(\ref{6.30}),(\ref{6.39c}) constrained
by (\ref{7.1}).

For this purpose one first observes that, according
to the mapping (\ref{2.0z}), the factor $e^{-S(\{\hat{1}\})}$ can be rewritten
as the partition function (PF) $\tilde{X}^{(a)}_{L^{D}}$ of the auxiliary
$D$-dimensional model. The latter is deduced from the intermediate
$D$-dimensional eigenvalue-system
(induced on $L^{D}$ lattice via the decomposition
(\ref{3.10d}),(\ref{3.10e}))
in the following way. Namely, in the plaquatte-factor defining the
latter eigenvalue-system, one is to substitute the link-variables
$U_{\rho}({\bf z})$ by the 'composite' field
\be
U_{\rho}({\bf z})\rightarrow
{\tilde{G}^{+}({\bf z})\tilde{G}({\bf z+\rho})}
\label{6.25v}
\ee
as it is predetermined by the pattern (\ref{2.0z}).
Consequently, the properly normalized PF of the induced gauge
theory can be represented as the ratio \cite{Dub1}
\be
X^{(in)}_{L^{D}}=X_{L^{D}}/X^{(a)}_{L^{D}}~,
\label{7.4v}
\ee
where $X_{L^{D}}$ and $X^{(a)}_{L^{D}}$ are the PFs (both normalized akin to
(\ref{7.3m})) associated to the intermediate $D$ dimensional
eigenvalue-system and the
auxiliary model defined through (\ref{6.25v}) respectively.

Next, the correspondence (\ref{2.10}) allows to express the large $N$ limit of
$X_{L^{D}}$ as the $L^{D}th$ power of the PF $X_{r}$ of the $k=1$ $D$-matrix
model (\ref{2.0})/(\ref{6.30}),(\ref{6.39c}). As we will prove in the end of
this subsection for the particular case of the latter model, the
$\{\lambda\}$-localization
(\ref{7.1}) results in the $power$-like asymptotics (provided $\int dU=1$) 
\be
\lim_{N\rightarrow{\infty}}
\lim_{\tilde{g}^{2}N\rightarrow{0}}
X_{r}(\{g_{r}\})=
[{B} ~\tilde{g}N^{\frac{1}{2}}]^{D N^{2}}~~,~~{B}>0~.
\label{7.4}
\ee
Being combined with the pattern (\ref{6.31}))/(\ref{6.39c}) of
$e^{-\tilde{S}^{(1)}_{r}(\{U\})}$, it ensures the complementary
$\{U_{\rho}\rightarrow{\hat{1}}\}$ localization (\ref{7.2b}) of the
fluctuations $A^{ab}_{\rho}$. Similarly to \cite{Dub1}, the pattern
(\ref{7.4}) is tantamount to the following large $N$ scaling-law (for each
particular $j$) 
\be
\lim_{N\rightarrow{\infty}}
\lim_{\tilde{g}^{2}N\rightarrow{0}} <\omega^{2}_{j}(\rho)>\sim{
O(\tilde{g}^{2}N)}~~\bmod (\ref{2.9})~,
\label{7.2}
\ee
where $\omega_{j}(\rho),~j=1,...,N,$ are the eigenvalues of
$U_{\rho}=\Omega_{\rho}~diag[e^{i\omega(\rho)}]~
\Omega_{\rho}^{+}$ entering the effective eigenvalue-theory (\ref{1.30}).
In turn, by the same token as in \cite{Dub1}, it ensures that in the
auxiliary model (\ref{6.25v}) the $SU(N)$ field $\tilde{G}({\bf z})$ is localized (modulo 
(\ref{2.9x})) in the vicinity of $\hat{1}$ so that
\be
\lim_{N\rightarrow{\infty}}
\lim_{\tilde{g}^{2}N\rightarrow{0}}
X^{(a)}_{L^{D}}(\{g_{r}\})=
[\tilde{B}~\tilde{g}N^{\frac{1}{2}}]^{L^{D} N^{2}}~~,~~\tilde{B}>0~.
\label{7.4vv}
\ee
Summarizing, we reproduce the purported asymptotics (\ref{7.4b}) of the
PF of the gauge theory induced from the $k=1$ GLR computable $D$-matrix model
(\ref{2.0})/(\ref{6.30}),(\ref{6.39c}).

In conclusion, let us demonstrate that in the $k=1$ $D$-matrix model
(\ref{2.0})/(\ref{6.30}),(\ref{6.39c}) the
large $N$ WC asymptotics (\ref{7.4}) is indeed valid.
To begin with, one readily obtains (from (\ref{6.31}),(\ref{6.39c})) for the
action of the latter model
\be
e^{-\tilde{S}^{(1)}_{r}(\{\hat{1}\})}=
\sum_{\{R^{(q)}_{\phi}\}}
e^{-\tilde{E}}~dim^{2}R_{+}
\prod_{p=1}^{2} L^{(D)}_{R_{+}|\{R^{(p)}_{\rho}\}}
\left(\otimes_{\mu=1}^{D}~L^{(D-1)}_{R^{(2)}_{\mu}|\{R_{\mu\nu}\}}\right),
\label{7.5}
\ee
where $\tilde{E}$ is defined by eq. (\ref{3.40}).
Thus, the asymptotics of the ratio
(\ref{7.4v}) is predetermined by the $\tilde{g}^{2}N$-scaling
of the factor
\be
[~L^{(D)}_{R_{+}|\{R^{(1)}_{\rho}\}}~dimR_{+}~\otimes_{\mu=1}^{D}
dimR^{(2)}_{\mu}~]^{-1}
\label{7.5bb}
\ee
responsible for the 'mismatch' between (\ref{6.39}) and
(\ref{6.31}) (where we have used that $R^{(2)}_{\mu}$ in (\ref{6.32}) can be
identified with $R_{\mu}$ of eq. (\ref{6.39b})). Next, recall that according
to eq. (\ref{3.43}), the irrep $R$ in eq. (\ref{1.2b}) represents the
irreps $\{R_{\rho}\}$ entering the GLR fusion-rules (\ref{6.3p}).
Owing to the pattern (\ref{6.31}),(\ref{6.39}),(\ref{6.39c}) of the
involved GLR fusion rules, the scaling-condition (\ref{7.1}) is valid for the
characteristic values of {\it all}
species $\{\lambda^{(0)}(\phi)\},~\phi\in{\{\mu\nu\},\{\rho\},+},$
parametrizing the SP irreps $\{R^{(0)}_{\phi}\}$
(on which the relevant large $N$ sums are localized).

Next, the dimension-formula
$\chi_{R(\{\lambda\})}(U)=det_{k,j}(e^{i\lambda_{k}\omega_{j}})/
det_{k,j}(e^{i(N-k)\omega_{j}})$ together with (\ref{7.1}) predetermines
that each $dimR_{\phi}$ contributes in the limit $N\rightarrow{\infty}$ with
the scaling-factor $[\tilde{g}N^{\frac{1}{2}}]^{-\frac{N^{2}}{2}}$.
Complementary, given (\ref{7.1}), the eigenvalues $\tilde{\omega}_{j}(\phi)$
of $\tilde{U}^{U(N)}_{\phi}$ (entering the definition (\ref{6.3}) of the relevant
GLR coefficients (\ref{7.5bb}) modified by the extension (\ref{3.9b}))
satisfy (\ref{7.2}) as in \cite{Dub1}.
Combining it with pattern (\ref{1.2d}) of the $U(N)$ Haar measure
($\int dU=1$),
one oncludes that the GLR coefficient of $Kth$ order scales in the large $N$
WC limit as $[dimR]^{K-1}
\sim{O([\tilde{g}^{2}N]^{-\frac{N^{2}(K-1)}{4}})}$.
Putting together all the scaling-factors inherent in (\ref{7.5bb}), one reproduces
(\ref{7.4}).

As a side remark, had we retained in $\tilde{X}_{r}$ of eq.
(\ref{6.39}) the $S(2n_{+})$-factor $K(R_{+})=
{\chi_{R_{+}}(\tilde{M}^{(1)}_{2n_{+}}\tilde{M}^{(2)}_{2n_{+}})}/
{d^{3}_{R_{+}}}$ while keeping
$E(\{R(\phi)\})$ ${\bf Z_{2}}$-invariant, there
would be $no$ way to adjust the parameters of the latter weight to set up
the $\{\lambda\}$-localization (\ref{7.1}).
Finally, we note also that the pattern of the large $N$ phase transitions
in the gauge theories induced from (\ref{2.0})/(\ref{6.30}),(\ref{6.39c})
can be analysed in essentially the same way as it is done
for the theories \cite{Dub1} induced from the eigenvalue-models
(\ref{1.35}),(\ref{1.36}).

\section{Conclusions.}

In this paper we have introduced the basic concepts of the nonabelian duality
transformation and applied them constructing a novel family
of solvable $D$-matrix models (defined via (\ref{6.1}))
graded by the rank $1\leq{k}\leq{D-1}$ of the manifest $[U(N)]^{\oplus k}$
conjugation-symmetry (\ref{2.0d}). The key-ingredients of the transformation
are the dual representation (\ref{3.6}) of
the $U(N)$ 1-link integral and the synthetic form (\ref{6.0}) of a generic
$D$-matrix $SU(N)$ or $U(N)$ system (\ref{2.0})/(\ref{6.20v}). Combining
these ingredients together, the partition function of {\it
any} matrix theory (\ref{2.0}) can be rewritten in terms of the $Tr_{4n_{+}}$
characters. The latter are the traces of the different $S(n_{\phi})$ tensors
(represented via (\ref{3.7})) which, being composed of the dual variables, are
embedded into the enveloping $S(4n_{+})$ space . The dual set consists of the
integer-valued $\{\lambda\}$
fields (parametrizing via (\ref{6.0b}) relevant irreps $\{R_{\phi}\}$)
that are complemented by the $\otimes_{n} S(n)$-valued degrees of freedom
(facilitating the fusion-rule algebra of the Young idempotents $C_{R_{\phi}}$
involved).

So far, the available solvable $D$-matrix models of the $1\leq{k}\leq{D-1}$
type are mainly associated to the situations \cite{Konts,Kaz&Mig} where an
application of the Itzykson-Zuber formula \cite{Itz&Zub}
transforms the model into some $k=D$ {\it eigenvalue}-theory of $q$
(hermitean or unitary) matrices. The proposed nonabelian duality suggests
the alternative 'mechanism' of the solvability realized for the subclass
(\ref{6.1}) of (\ref{2.0}). Here the underlying reason is the hidden symmetry
of the action which becomes manifest after reformulation in terms of the
dual variables. It is this somewhat unconventional symmetry which
predetermines that the involved Young idempotents satisfy the {\it simplest}
pattern (\ref{6.2}) of the fusion-rules encoded by the GLR coefficients
(\ref{6.3}). In turn, the GLR pattern allows to map the associated
$1\leq{k}\leq{D-1}$ systems (\ref{2.0}),(\ref{6.1}) onto the 
$D$-matrix eigenvalue-models (\ref{1.35}),(\ref{1.36}) endowed with the
largest possible $k=D$ conjugation-symmetry (\ref{2.0d}).

The latter $k=D$ eigenvalue-models, being solvable in the limit
$N\rightarrow{\infty}$, has been recently proposed \cite{Dub1} to reproduce
the large $N$ free energy (FE) of the associated lattice gauge theory in $D$
dimensions. Generalizing the algorithm of \cite{Dub1}, the prescription is
developed to reconstruct the gauge theory with the FE
$-ln[\tilde{X}_{L^{D}}]$ equal (modulo the $L^{D}$-volume factor (\ref{2.10}))
to the FE $-ln[\tilde{X}_{r}]$ of a given $1\leq{k}\leq{D-1}$
system (\ref{2.0}),(\ref{6.20v}). The new algorithm is applicable to
any hypothetical solvable $D$-matrix model (i.e. is {\it not} necessarily restricted to
the subvariety (\ref{6.1})) consistent with the $[{Z_{N}}]^{D}$-invariance
(\ref{2.9}).

As well as in \cite{Dub1}, we address the question of the continuum limit in
the gauge systems induced from (\ref{2.0}),(\ref{6.1}).
To clarify this issue, we choose the judiciously constructed  
$k=1$ family (\ref{6.30}) {\it constrained} by (\ref{6.39c}). To be even more
specific, the 1-matrix model of the representation (\ref{1.2b}) is selected
in the simplest form
(\ref{4.1}). Given this choice, we prove that the proposed in \cite{Dub1}
scaling-condition (\ref{7.1}) (imposed on the effective 1-matrix system
(\ref{1.2b})) does ensure the required localization
$\{U_{\rho}({\bf z})\rightarrow{\hat{1}}\}$ of the link-variables in the
associated induced gauge theories.

The major motivation for this project is to develope the formalism
which makes accessible the Gauge String representation (of strongly coupled
gauge theories) yielding the $D>2$ extension of the $D=2$ construction due to
Gross and Taylor \cite{Gr&Tayl}. One observes that the proposed approach deals
with the structures which are already very similar in nature to those of
\cite{Gr&Tayl}. The precise reformulation of the amplitudes (in a lattice
$YM$ system), in compliance with the pattern defining the data of the properly
associated (generalized) branched covering spaces, will be given in the
forthcoming paper \cite{Dub2}. Here we announce only one
important novel feature of the $D>2$ stringy pattern that is not present in
the $D=2$ construction. Compared to the latter case, in $D>2$ one is forced
to introduce certain generalizations of the canonical branched covering
spaces. This is foreshadowed algebraically by the necessity to embed various
$S(n_{\phi})$ operators
(acting in {\it different} subspaces) into the common enveloping space in
a manner similar to what we have done in Sections 2 and 3.

\begin{center}
{\bf Acknowledgements.}
\end{center}

This project was started when the author was the
NATO/NSERC Fellow
at University of British Columbia, and I would like to thank all the stuff and
especially Gordon Semenoff for hospitality.

\app{$SU(N)$/$\bigotimes_{n} S(n)$ complementarity.}

To facilitate the nonabelian Duality transformation on the lattice, one needs
a piece of the formalism which we now focus on. Let us start with the basic
facts about the action of the $GL(N)$ (which can be further restricted to
$U(N)$ or $SU(N)$) and the symmetric groups on the
tensor spaces associated to the structures introduced in the Section 2.

Recall that $GL(N)$ is the group of
the automorphisms of nondegenerate ($detV\neq{0}$) complex $N\times N$
matrices $V_{i}^{j}$. Given any basis $\{ |i>,~i=1,2,...,N\}$ on a
$N$-dimensional vector space $X_{N}$, the fundamental
matrix representation of $GL(N)$ is defined canonically as
$\hat{V}|i>=V_{i}^{j}|j>$. Given $X_{N}$, one introduces the basis
$|i>^{\oplus n}=|i_{1}>\otimes |i_{2}>\otimes ...|i_{n}>$ for the direct
product space
$X_{N}^{\oplus n}$. The elements of $GL(N)$ act on $X_{N}^{\oplus n}$
according to the standard rule $\hat{V}\otimes_{p=1}^{n}|i_{p}>=
D(V)_{\{i^{\oplus n}\}}^{\{j^{\oplus n}\}}|j>^{\oplus n}$, where
conventionally $D(V)_{\{i^{\oplus n}\}}^{\{j^{\oplus n}\}}=
(V)_{i_{1}}^{j_{1}}(V)_{i_{2}}^{j_{2}}...  (V)_{i_{n}}^{j_{n}}
\equiv{V^{\oplus n}}$.

Next, the representation theory proves (see e.g. \cite{Gr-in-phys}) that the
symmetric group $S(n)$ is the most general group of transformations commuting
with the elements $V^{\oplus n}$ of the $GL(N)$ on the $X_{N}^{\oplus n}$.
The elements of the $S(n)$-group are represented by the linear tranformation
\be
\sigma|i>^{\oplus n}=|i_{\sigma^{-1}(1)}>\otimes |i_{\sigma^{-1}(2)}>\otimes
... |i_{\sigma^{-1}(n)}>=
D(\sigma)_{\{i^{\oplus n}\}}^{\{j^{\oplus n}\}}|j>^{\oplus n}
\label{A.3}
\ee
where $D(\sigma)_{\{i^{\oplus n}\}}^{\{j^{\oplus n}\}}$ is given
by eq. (\ref{3.7}), with the basic property being the commutativity
\be
[D(\sigma),D(V)]=0~~,~~\forall \sigma\in{S(n)},~\forall V\in{G}~.
\label{A.4}
\ee
The representation of the $S(n)$-algebra elements is deduced from the
group-representation (\ref{3.7}) by linearity. Remark that the unity element
$\hat{1}_{[n]}$ of the $S(n)$ group is represented by the 'trivial'
permutation (\ref{A.3}) $\hat{1}_{[n]}$: $\sigma(k)=k,~k=1,...,n$.

The central operation, we will employ, is the decomposion of the direct
product
space $V^{\oplus n}$ (or, equally, $(V^{+})^{\oplus n}$) into the irreps of
the Lie group or, $dually$, into the irreps of $S(n)$.
Recall that in the $S(n)$ group the irreps are labelled \cite{Gr-in-phys} by a
set of $k$ nonnegative, nonincreasing integers
$\{n_{i};~n_{1}\geq{n_{2}}\geq{...}\geq{n_{k}}\geq0\}$
constrained by the single condition $\bigoplus_{i=1}^{k} n_{i}=n$
and visualized as the Young tableau $Y_{n}$ with $n$ boxes. The $U(N)$ or
$GL(N)$ irreps can be parametrized in a similar fashion by a set of $N$
integers $\{n_{i};~n_{1}\geq{n_{2}}\geq{}...\geq{n_{N}}\}$ that is related to
the alternative classification (\ref{2.3}) identifying $n_{i}=\lambda_{i}-N+i$.
When all $n_{i},~i=1,...,N,$ are nonnegative (nonpositive), the associated
$U(N)$-characters are expressed in terms of the $V$- ($V^{+}$-) tensors only.
The corresponding '$(anti)chiral$' sector of $U(N)$-irreps can be visualized
\cite{Gr-in-phys} by the $U(N)$ Young tableaus $Y_{n}^{(N)}$ containing
$n=\sum_{i=1}^{N} n_{i}$ boxes distributed in not more than $N$ rows.
In the $SU(N)$ case, the complete set of irreps is labelled by the $SU(N)$
Young tableaus $Y_{n}^{(N)}$ containing not more than $N-1$ rows, i.e.
$n_{N}=0$.

The required decomposition of $V^{\oplus n}$ (or $(V^{+})^{\oplus n}$) can
be canonically generated by the $Y_{n}^{(N)}$-$subset$ of the Young
projectors $\{P_{R}\}$ (defined by eq. (\ref{3.9})) that belong to the center
of the $S(n)$ algebra: $[P_{R},\sigma]=0,~\forall{\sigma\in{S(n)}}$,
$\forall{R\in{Y_{n}}}$.
The tensor product $V^{\oplus n}$ is mapped \cite{Gr-in-phys} by an admissible
projector $P_{R},~R\in{Y_{n}^{(N)}}$ onto the $dimR$ copies of the $S(n)$
irrep $\tilde{T}_R$ or
equivalently onto the $d_{R}$ copies of the Lie group irrep $T_{R}$
\be
D(P_{R})V^{\oplus n}=(\tilde{T}_{R})^{\oplus dimR}=(T_{R})^{\oplus d_{R}}=
{\tilde{T}_{R}\otimes T_{R}}~~,~~R\in{Y_{n}^{(N)}}~,
\label{A.8}
\ee
where $d_{R},dimR$ stand respectively for the dimensions of $S(n)$- 
and the
chiral $GL(N)$- (or,
equivalently, $U(N)$-) irreps $R$ respectively. In the case of
$G=SU(N)$, one
obtains in this way all the irreducible representations. For $G=U(N)$, the
space $V^{\oplus n}$ contains irreps included into the single chiral sector
of $U(N)$-irreps parametrized by the Young tableaus $Y_{n}^{(N)}$.

Combining equation (\ref{A.8}) with the defining properties of $\{P_{R}\}$
\be
P_{R_{1}}P_{R_{2}}=\delta_{R_{1},R_{2}}P_{R_{1}}~~ ,~~\sum_{R\in{Y_{n}}}
P_{R}=1
\label{A.9}
\ee
we deduce one of the central results of the representation theory, the
so-called Schur-Weyl duality (see e.g. \cite{Moore})
\be
V^{\oplus n}=\sum_{R\in{Y_{n}^{(N)}}}{\tilde{T}_{R}\otimes T_{R}}
\label{A.10}
\ee
which formalizes the complementary roles of the Lie and symmetric groups.

Employing eqs. (\ref{A.8}),(\ref{A.10}), the formulas relevant for the Duality
transformation can be represented in the concise algebraic form. First,
taking trace of the Schur-Weyl decomposition (\ref{A.10}) and using the
completeness condition (\ref{A.9}) for $\{P_{R}\}$, one deduces
(see e.g. \cite{Moore}) the second Frobenius formula
\be
\Upsilon_{[\sigma]}(V)=Tr_{n} [D(\sigma)V^{\oplus n}]=
\sum_{R\in{Y_{n}^{(N)}}} \chi_{R}(\sigma)\chi_{R}(V)
\label{A.12}
\ee
or its modification  
$Tr_{n} [D(P_{R})D(\sigma)V^{\oplus n}]=\chi_{R}(\sigma)\chi_{R}(V)$
following from eq. (\ref{A.8}). Similarily, the first Frobenius formula
(\ref{3.10}) for $\chi_{R}(V)$ can be obtained multiplying eq. (\ref{A.10})
by $D(\sigma)$ and taking the trace as previously.

\app{The Dual form of the 1-link integral.}

In this Appendix, we derive the dual representation (\ref{3.6}) of the 1-link
integral (\ref{3.5}). To compute
$M^{G}(n,m)^{p_{1}...q_{m}}_{j_{1}...l_{m}}$ of eq. (\ref{3.5})
for a particular Lie group $G$, the starting
idea \cite{Samuel,Dr&Zub} is to differentiate the simple generating function
\be
F^{G}_{n,m}(A,B)=\int{dV}~(Tr[AV])^{n}~(Tr[BV^{+}])^{m}~
\label{B.1}
\ee
with respect to $A,B\in{G}$.

\sapp{The $S(n)$-form of the $U(N)$ formula.}

In the $U(N)$-case the invariance (\ref{2.9}) of the integral (\ref{B.1})
under the $U(1)$-subgroup of $U(N)=(SU(N)\otimes U(1))/Z_{N}$ ensures that
it doesn't vanish only for $n=m$. For $F_{n,n}(A,B),~A,B\in{U(N)}$, one
derives \cite{Samuel,Dr&Zub}
\be
F^{U(N)}_{n,n}(A,B)=
{\sum_{R\in{Y_{n}^{(N)}}}}{\frac{d_{R}^{2}}{dimR}}\chi_{R}(AB)~,
\label{B.2}
\ee
where the second Frobenius formula (\ref{A.12}) and the standard orthogonality
condition of the characters have been employed.
Applying to $F_{n,n}(A,B)$ the operator
$(n!)^{-2}\prod_{k=1}^{n} { {\partial^{2}}/
{\partial A_{p_{k}}^{j_{k}}\partial B_{q_{k}}^{l_{k}}}}$
(where $A,B$ in the r.h.s. of eq. (\ref{B.2}) can be extended to $GL(N)$), 
one obtains for the $U(N)$ 1-link integral (\ref{3.5})
\be
{\frac{1}{(n!)^{2}}}
{\sum\limits_{R\in{Y_{n}^{(N)}}}}{\frac{d_{R}^{2}}{dimR}}
\sum\limits_{\rho,\sigma\in{S(n)}}
D(C_{R})_{j_{\rho(1)}j_{\rho(2)}...j_{\rho(n)}}^
{q_{\sigma(1)}q_{\sigma(2)}...q_{\sigma(n)}}~
\delta^{p_{\rho(1)}}_{l_{\sigma(1)}} \delta^{p_{\rho(2)}}_{l_{\sigma(2)}}...
\delta^{p_{\rho(n)}}_{l_{\sigma(n)}}.
\label{B.4}
\ee
Introducing $\delta=\rho{\sigma}^{-1}\in{S(n)}$, we arrive at the concise
representation given by the eq. (\ref{3.6}), with the $S(n)$-tensors
$D(\delta)$ and $\Lambda^{(-1)}_{n}$ being defined by eqs. (\ref{3.7}) and
(\ref{3.8}) respectively. Note that, in $\Lambda^{(-1)}_{n}$ the sum over the
$U(N)$ G-irreps $R\in{Y_{n}^{(N)}}$ is restricted to the single chiral
sector parametrized by the $U(N)$ Young tableaus $Y_{n}^{(N)}$ (containing not
more than $N$ rows). Let us remark
also that eq. (\ref{3.6}) is complementary to already existing representations
\cite{Samuel,Dr&Zub},\cite{Gr&Tayl} of (\ref{3.5}) which are not suitable for
our present purposes.

\sapp{The $S(2n)$-form of the $U(N)$ formula.}

The integration formula (\ref{3.6}) can be represented in terms of the
elements of the $S(2n)$-algebra 'enveloping' $S(n)\otimes S(n)$. Employing the
ordered link-basis $|I_{2n}>=|I^{(+)}_{n}>|I^{(-)}_{n}>$ (with
$|I^{(\pm)}_{n}>=|i_{\pm}>^{\oplus n}$ akin to (\ref{5.13b})),
one rewrites eq. (\ref{3.6}) in the $S(2n)$-form of eqs.
(\ref{5.11}),(\ref{5.12}). By construction, the 'chiral', $|I^{(+)}_{n}>$, and
the 'antichiral', $|I^{(-)}_{n}>$, sectors are associated respectively to
the first and to the second $S(n)$-subblocks of (\ref{5.11}),(\ref{5.12}). As
for the operator $\Phi_{2n}\in{S(2n)}$, comparing (\ref{3.6}) and (\ref{5.11})
one deduces for its explicit form
\be
D(\Phi_{2n})^{\{j^{\oplus 2n}\}}_{\{i^{\oplus 2n}\}}=
\delta^{j_{1}}_{i_{n+1}}\delta^{j_{2}}_{i_{n+2}}...\delta^{j_{n}}_{i_{2n}}~
\delta^{j_{n+1}}_{i_{1}}\delta^{j_{n+2}}_{i_{2}}...\delta^{j_{2n}}_{i_{n}}
\label{B.20}
\ee
which in turn can be concisely represented in the alternative ordered $S(2n)$-basis
$|\tilde{I}_{2n}>=(|i_{+}>|i_{-}>)^{\oplus n}$ as the outer product
\be
D(\Phi_{2n})=D({(c_{2})}^{\oplus n})\in{S(2n)}
\label{B.22}
\ee
of the 2-cycle permutations $c_{2}\in{C_{2}}$,
$c_{2}:\{12\}\rightarrow{\{21\}}$, where each individual $c_{2}\in{S(2)}$ acts
on the $(|i_{+}>|i_{-}>)$-subspace of $|\tilde{I}_{2n}>$.

Finally, we mention the commutation-rules 
\be
[~\Phi_{2n}~,~\Gamma(2n)~]=
[~\Gamma(2n)~,~(\Lambda^{(m)}_{n}\otimes{\hat{1}_{[n]}})~]=0~,
\label{B.23}
\ee
where in the second case the above $|I_{2n}>$-basis is employed.
In particular, it makes the relative order of the operators in the product
(\ref{5.11}) immaterial. Also, let us include the following useful identities
$$\Gamma(2n)\cdot (E^{(+)}_{n}\otimes E^{(-)}_{n})=
(E^{(-)}_{n}\otimes E^{(+)}_{n})\cdot \Gamma(2n)~;$$
\be
\Phi_{2n}\cdot (E^{(+)}_{n}\otimes E^{(-)}_{n})=
(E^{(-)}_{n}\otimes E^{(+)}_{n})\cdot \Phi_{2n}~,
\label{B.24}
\ee
valid for $\forall{E^{(\pm)}_{n}}\in{S(n)}$ where again the $|I_{2n}>$-basis
is implied.

\sapp{$V^{SU(N)}\rightarrow{V^{U(N)}}$ extension for 
$F^{SU(N)}_{n,n}(A,B)$.}

Let us prove that the diagonal $SU(N)$ moments $F^{SU(N)}_{n,n}(A,B)$, defined
by eq. (\ref{B.1}), are invariant under the substitution of the $SU(N)$
link-variables by the $U(N)=[SU(N)\otimes U(1)]/{\bf Z_{N}}$ ones
\be
V^{SU(N)}\rightarrow{(V^{SU(N)}\otimes V^{U(1)})/{\bf Z_{N}}}=V^{U(N)}~~,~~
dV^{SU(N)}\rightarrow{dV^{U(N)}}~,
\label{B.9}
\ee
that is tantamount to eq. (\ref{3.9b}). Actually, the identity (\ref{3.9b})
for $n<N$ directly follows from equivalence \cite{Dr&Zub} of any polynomial  
$U(N)$ representation with $n<N$ to the corresponding $SU(N)$ one. But for
$n\geq{N}$ one needs a more refined consideration (valid for any $n$) we
now focus on.

First, we note that the extra $U(1)$-factor in eq. (\ref{B.9})
matches with the auxiliary $U(1)$-invariance (\ref{2.9}) which is implicit in
the diagonal 'moments' $F^{SU(N)}_{n,n}(A,B)$ (while nondiagonal
$n\neq{m}$ $SU(N)$ integrals (\ref{3.5}) are only ${\bf Z_{N}}$-invariant).
To promote (within $F^{SU(N)}_{n,n}(A,B)$) the ${\bf Z_{N}}$ center-subgroup
(\ref{2.9}) into $U(1)$, one is to multiply each $V^{SU(N)}$ by the auxiliary
factor $V^{U(1)}$ and then integrate $dV^{U(1)}$ with the $U(1)$ Haar
measure normalized to unity.
The resulting pattern is to be confronted with
the factorized representation of the $U(N)$ measure 
\be
\int_{U(N)} dV^{U(N)}...\rightarrow{\int_{U(1)} dV^{U(1)}
\int_{SU(N)} dV^{SU(N)}}...
\label{4.10}
\ee
that will be derived below. Identifying the $U(1)$ sector in eq.
(\ref{4.10}) with the averaging over the auxiliary $U(1)$-transformation
(that leaves diagonal integrals $F^{SU(N)}_{n,n}(A,B)$ invariant), we justify
the required formula (\ref{3.9b}).

To obtain eq. (\ref{4.10}), recall first that the explicit expressions
\cite{Dr&Zub} for the $U(N)$ and $SU(N)$ measures (after decomposition
$V=\Omega~diag[e^{i\omega}]~\Omega^{+}$) are given respectively by eq.
(\ref{1.2d}) and 
\be
\int dV^{SU(N)}...=\int_{-\pi}^{+\pi}\prod_{k=1}^{N} \frac{d\omega_{k}}{2\pi}
~\delta^{(2\pi)}(\omega_{+})~
|\Delta(\{\omega_{p}\})|^{2}~d\Omega...~,
\label{B.11}
\ee
where $\omega_{+}=\sum\limits_{k=1}^{N} \omega_{k}$, and 
$\delta^{(2\pi)}(\phi)=2\pi \sum_{n\in{\bf Z}}\delta(\phi-2\pi n)$ is the
periodic $\delta$-function.

Next, the factorized form (\ref{4.10}) of the $U(N)$ measure (\ref{1.2d}) is
predetermined by the decomposition $V^{U(N)}=
(V^{U(N)}/det[V^{U(N)}]^{1/N})\otimes det[V^{U(N)}]^{1/N}$ that allows to
identify
\be
(V^{U(N)}/det[V^{U(N)}]^{1/N})
\cong{V^{SU(N)}(\{\omega^{SU(N)}_{k}\},\Omega)}~,
\label{B.13}
\ee
\be
V^{U(1)}\cong{det[V^{U(N)}]^{1/N}}=e^{i\omega^{U(N)}_{+}/N}~,
\label{B.14}
\ee
where $\omega^{U(N)}_{+}/N\in{[-\pi,+\pi]~\bmod 2\pi}$. As a result, in eq.
(\ref{4.10}) $V^{U(1)}\equiv{V^{U(1)}(\omega_{+}/N)}$, 
$\int dV^{U(1)}=\int_{-\pi}^{+\pi} d(\omega_{+}/N)/2\pi$, while
\be
\omega^{SU(N)}_{k}=
\omega^{U(N)}_{k}-\frac{\omega^{U(N)}_{+}}{N}~~;~~
\omega^{SU(N)}_{+}=\sum_{k=1}^{N} \omega^{SU(N)}_{k}=0 \bmod 2\pi.
\label{B.15}
\ee
In order to convert the decomposition (\ref{B.13}),(\ref{B.14}) into
(\ref{4.10}), one is to change
variables in eq. (\ref{1.2d}) going over from $\{\omega^{U(N)}_{k}\}$ to the
overcomplete set $\{\omega^{SU(N)}_{k};~k=1,...,N\}\otimes
(\omega^{U(N)}_{+}/N)$
entering eq. (\ref{B.15}). First,  
$\Delta(\{\omega^{U(N)}_{k}\})=\Delta(\{\omega^{SU(N)}_{k}\})$.
Second, the constraint $\omega^{SU(N)}_{+}=0 \bmod 2\pi$
for the $\{\omega^{SU(N)}_{k}\}$-set can be imposed by the
$\delta^{(2\pi)}$-function (reminiscent of the $SU(N)$ measure (\ref{B.11})),
and no additional Jacobian is needed. Summarizing, we arrive at the identity
\be
dV^{U(N)}(\{\omega^{U(N)}_{k}\},\Omega)=
dV^{U(1)}(\omega^{U(N)}_{+}/N)dV^{SU(N)}(\{\omega^{SU(N)}_{k}\},\Omega)~,
\label{B.12}
\ee
\be
\omega^{SU(N)}_{k}\in
{[-\pi+\frac{\omega^{U(N)}_{+}}{N}~,~\pi+\frac{\omega^{U(N)}_{+}}{N}~]}~
\bmod 2\pi~,~
k=1,...,N,
\label{B.16}
\ee
where $\int dV^{U(N)}$ and $\int dV^{SU(N)}$ are defined by eqs. (\ref{1.2d})
and (\ref{B.11}) respectively. Evaluating the diagonal $SU(N)$ 1-link
integral $F^{SU(N)}_{n,n}(A,B)$, one observes that the overall
$(\omega^{U(N)}_{+}/N)$-shift (\ref{B.16}) of the
$\{\omega^{SU(N)}_{k}\}$-variables $doesn't$ affect the result. Altogether,
it transforms the equality (\ref{B.12}) into the announced decomposition
(\ref{4.10}) of the $U(N)$ measure.

Finally, the invariance (\ref{2.9}) of the diagonal $SU(N)$ 1-link integral
(\ref{B.1}) under the extension of the measure
\be
V^{SU(N)}(\{\omega^{SU(N)}_{k}\},\Omega)\rightarrow{
V^{U(1)}(\omega^{U(N)}_{+}/N)~
V^{SU(N)}(\{\omega^{SU(N)}_{k}\},\Omega)}
\label{B.17}
\ee
allows to identify $F^{SU(N)}_{n,n}(A,B)$ with the $F^{U(N)}_{n,n}(A,B)$.
For this purpose, one is to integrate
$\int_{-\pi}^{+\pi} d(\omega_{+}/N)/2\pi$
that trades (\ref{B.11})
for the factorized pattern (\ref{4.10}) of the $U(N)$ measure. This completes
the proof of the identity (\ref{3.9b}).

\app{Fusion rules of Young idempotents.}

In this Appendix we derive the identities relating Young idempotent
$C_{R_{+}}\in{S(n_{+})},~n_{+}=\sum\limits_{k=1}^{p} n_{k}$,
with the corresponding direct product $\otimes_{k=1}^{p} C_{R_{k}}$. Let us
start with the following observation. The outer product of two idempotents
$C_{R_{1}} \otimes{C_{R_{2}}}\in{S(n_{1})\otimes{S(n_{2})}}$, being embedded
into $S(n_{1}+n_{2})$, $ceases$ to be a composition $\phi_{\alpha}$ of the
$S(n_{1}+n_{2})$ Young idempotents defined as 
\be
\phi_{\alpha}=\sum_{\sigma\in{S(n_{+})}} \alpha([\sigma])~\sigma=
\sum_{R_{+}\in{Y_{n_{+}}^{(N)}}}\alpha_{R_{n_{+}}}~C_{R_{n_{+}}}~,
\label{D.1}
\ee
where $\alpha([t\sigma t^{-1}])=\alpha([\sigma])$, for
$\forall{t\in{S(n_{1}+n_{2})}}$. Indeed, (for $R_{k}\in{Y_{n_{k}}^{(N)}}$)
the product $(C_{R_{2}} \otimes{C_{R_{1}}})V^{\oplus (n_{1}+n_{2})}$, being a
Lie group representation
\be
Tr_{n_{+}}[(\otimes_{k=1}^{p} C_{R_{k}})
V^{\oplus n_{+}}]=\otimes_{k=1}^{p}\chi_{R_{k}}(V)~;~~
R_{k}\in{Y_{n_{k}}^{(N)}}~,~
n_{+}=\sum_{k=1}^{p} n_{k},
\label{D.2}
\ee
does $not$ generate a $S(n_{1}+n_{2})$-representation. To preserve the
Schur-Weyl duality (\ref{A.8}), we introduce the '$twisted$' deformation
\be
C_{R_{1}} \otimes{C_{R_{2}}}\rightarrow
{\frac{1}{(n_{1}+n_{2})!}\sum\limits_{\delta\in{S(n_{1}+n_{2})}}
[~\delta~ (C_{R_{1}}\otimes{C_{R_{2}}})~ \delta^{-1}~]}~.
\label{D.2bb}
\ee
 By construction, the 'twisted' product (\ref{D.2bb}) commutes with
 $\forall{t\in{S(n_{1}+n_{2})}}$ and thus
belongs to the center (\ref{D.1}) of the $S(n_{1}+n_{2})$-algebra. More
generally, given $V\in{U(N)}$ we arrive at the pattern
(\ref{6.2}) of the fusion rules of the $S(n_{\psi})$-valued Young
idempotents. By the same token, one arrives at the inverse of the identity
(\ref{6.2})
\be
C_{R_{+}}=\bigoplus_{\{R_{k}\in{Y_{n_{k}}}\}}L^{(p)}_{\{R_{k}\}|R_{+}}
\sum\limits_{\delta\in{S(n_{+})}}
\frac{[~\delta~ (\otimes_{k=1}^{p} C_{R_{k}})~ \delta^{-1}~]}{(n_{+})!}
~,~~
R_{+}\in{Y_{n_{+}}}~,
\label{D.5}
\ee
expressed, strictly speaking, in terms of some other set of the GLR
coefficients $L^{(p)}_{R_{1}...R_{p}|R_{+}}
\equiv{L^{(p)}_{\{R_{k}\}|R_{+}}}$ of $pth$ order.

Next, in the framework of the duality transformation the fusion rules
(\ref{6.2}) and (\ref{D.5}) are realized in the tensor representation
(\ref{3.7}) and enter only {\it inside} the associated traces $Tr_{m}$. 
Consequently, according to the second Frobenius formula (\ref{A.12}),
it effectively eliminates the contribution of those $S(n_{\psi})$-irreps
$R_{\psi} \in{Y_{n_{\psi}}}$ which do $not$ correspond to a $U(N)$ irrep
(i.e. do not belong to $Y_{n_{\psi}}^{(N)}$). Therefore, we confine our
attention to the GLR coefficients $L^{(p)}_{R_{+}|\{R_{k}\}},~
L^{(p)}_{\{R_{k}\}|R_{+}}$ with {\it all} $R_{\psi}\in{Y_{n_{\psi}}^{(N)}}$.
Upon a reflection, this subset coincides with the coefficients in the
associated via (\ref{D.2}) decomposition of the $U(N)$ characters (with the
irreps restricted to the chiral sector), i.e. assume the integral form
\be
L^{(p)}_{R_{+}|\{R_{k}\}}=L^{(p)}_{\{R_{k}\}|R_{+}}=
\int_{U(N)} dV \chi_{R_{+}}(V^{+})\otimes_{k=1}^{p}\chi_{R_{k}}(V)~,
\label{6.2vxx}
\ee
where $n_{+}=\sum_{k=1}^{p} n_{k}$.
To prove this statement, one first applies both sides
of (\ref{6.2}) (or (\ref{D.5})) to $V^{\oplus n_{+}}\in{U(N)}$
and then takes the overall trace $Tr_{n_{+}}$. Employing the
commutativity (\ref{A.4}), we get rid of the sum over $\delta\in{S(n_{+})}$
and use the identity (\ref{D.2}) which altogether results in (\ref{6.2vxx}).
Finally, let us remark that a generic coefficient $L^{(p)}_{R_{+}|\{R_{k}\}}
,~\forall{R}_{k}\in{Y_{n_{k}}^{(N)}},$ can be represented as a
combination of the 'elementary', $p=2$ Littlewood-Richardson (LR)
coefficients $L^{(2)}_{R_{+}|R_{1}R_{2}}$ \cite{Gr-in-phys} (entering the
two-character fusion rules).

\end{document}